\documentclass[12pt]{article}
\usepackage{amssymb,amsmath,epsfig,cite,xcolor}
 \allowdisplaybreaks

\begin{document}
\title{\bf Orthogonal splitting of the Riemann curvature tensor and its implications
in modeling compact stellar structures}
\author{A. Rehman${^1}$ \thanks{v31763@umt.edu.pk}~,
Tayyab Naseer$^{2,3}$
\thanks{tayyab.naseer@math.uol.edu.pk; tayyabnaseer48@yahoo.com}~,
Nazek Alessa$^4$ \thanks{naalessa@pnu.edu.sa}~ \\
and Abdel-Haleem Abdel-Aty$^5$ \thanks{amabdelaty@ub.edu.sa}\\
${^1}$Department of Mathematics, University of Management and
Technology,\\
Johar Town Campus, Lahore-54782, Pakistan.\\
$^2$Department of Mathematics and Statistics, The University of Lahore,\\
1-KM Defence Road Lahore-54000, Pakistan.\\
$^3$Research Center of Astrophysics and Cosmology, Khazar University, \\
Baku, AZ1096, 41 Mehseti Street, Azerbaijan.\\
$^4$Department of Mathematical Sciences, College of Science,\\
Princess Nourah bint Abdulrahman University,\\ P.O. Box 84428,
Riyadh 11671, Saudi Arabia.\\
$^5$Department of Physics, College of Sciences, \\University of
Bisha, Bisha 61922, Saudi Arabia.}

\date{}

\maketitle
\begin{abstract}
Although the interpretation of complexity in extended theories of
gravity is available in the literature, its illustration in
$f(R,L_{m},\mathcal{T})$ theory is still ambiguous. The orthogonal
decomposition of the Riemann tensor results in the emergence of
complexity factor as recently proposed by Herrera
\cite{herrera2018new}. We initiate the analysis by contemplating the
interior spacetime as a static spherical anisotropic composition
under the presence of charge. The modified field equations are
derived along with the establishment of association between the
curvature and conformal tensors that have significant relevance in
evaluating complexity of the system. Furthermore, the generalized
expressions for two different masses are calculated, and their link
with conformal tensor is also analyzed. Moreover, we develop a
particular relation between predetermined quantities and evaluate
the complexity in terms of a certain scalar $Y_{TF}$. Several
interior solutions admitting vanishing complexity are also
determined. Interestingly, compact objects having anisotropic matter
configuration along with the energy density inhomogeneity possess
maximum complexity. It is concluded that the spherical distribution
of matter might not manifest complexity or admitting minimal value
of this factor in the framework of $f(R,L_{m},\mathcal{T})$ theory
due to the appearance of dark source terms.
\end{abstract}

\section{Introduction}

Efforts have been initiated in recent years to develop an accurate
methodology for assessing complexity in numerous scientific realms
\cite{calbet2001tendency,sanudo2008statistical}. Nevertheless,
despite these attempts, there is still no consensus on its precise
description. The molecules in the case of perfect crystal are
uniformly aligned resulting in insignificant information. However,
the structure of an ideal gas provides much more information because
of its inconsistent composition of atoms. Due to the simplest
configuration, the aforementioned perfect crystal and ideal gas
frameworks claim no complexity. The higher amount of information and
structural layout in these models suggests the need for additional
specifications in the description of complexity. Earlier approaches
to define complexity relied on entropy and information related to
the matter composition. L{\'o}pez-Ruiz et al.
\cite{lopez1995statistical} suggested the novel description of
complexity having dependence on disequilibrium of the system that is
maximum in perfect crystal whereas turns to be zero in the scenario
of an ideal gas. This particular notion explained the equivalence of
complexity for both frameworks. It is important to mention that the
description of complexity proposed by L{\'o}pez-Ruiz has already
been considered for the elaboration of self-gravitating compact
objects \cite{sanudo2009complexity,chatzisavvas2009complexity}. The
pressure exerted by the fluid dispersion is an important factor in
evaluating the fundamental features and development of cosmic
formations. Consequently, we can say that the inclusion of pressure
in the notion of complexity is also of significant relevance.

After contemplating the basic formalism of general relativity (GR),
Herrera \cite{herrera2018new} suggested an entirely new idea for
interpreting complexity, according to which the combination of
non-uniform density with the anisotropic pressure illustrates the
intricacy associated with the self-gravitational compact objects.
The trace-free part $Y_{TF}$ incorporating the above-mentioned
variables is resulted after the orthogonal decomposition of the
electric part of the Riemann curvature tensor. This scalar quantity
is an in determining the complexity of a system and is termed as
complexity factor (CF). The quantity $Y_{TF}$ disappears in following scenarios:\\
$\bullet$ The fluid distribution is isotropic and possesses uniform energy density.\\
$\bullet$ The two terms comprising the non-uniform density and the anisotropic stresses cancel each other.\\
Herrera et al. \cite{herrera2018definition} further explained the
essence of complexity related to the static composition of fluid.
They established the minimal criterion of complexity and concluded
that the distribution of fluid is geodesic yielding numerous
solutions. The relevance of complexity in different physical
compositions in the background of the axial fluid distribution can
be seen in \cite{herrera2019complexity}. Herrera et al.
\cite{herrera2020quasi} considered the concept of complexity for
establishing the non-static spherical composition after assuming the
fluid possessing dissipation and non-dissipation. They determined
relations between areal radius and areal velocity along with some
solutions whose validity has also been examined in explaining the
accelerated expansion of the universe. In the context of this newly
developed notion, Contreras and Fuenmayor
\cite{contreras2021gravitational} assessed the self-gravitating
spherical celestial composition through the mechanism of gravitating
cracking. The effects of fluctuations in decoupling parameters on
pressure were analyzed along with the density of celestial objects.
Herrera et al. \cite{herrera2021hyperbolically} analyzed the effect
of different structure scalars determined from the orthogonal
decomposition of the curvature tensor and demonstrated the idea of
complexity associated with the hyperbolically symmetric composition of
structure. The analysis of complexity under various gravitational
frameworks is available in the literature
\cite{yousaf2022study,bhatti2022novel,yousaf2020new,naseer2024implications,bhatti2025complexity,
naseer2024extending,sharif2024study,nasir2023influence,aa,ab,ac,ad,ae,af,ag,ah,ai,ba,bb,bc,bd,be,bf,bg,bh,bi,
bj,bk,bl,bm}.

The existence of electromagnetic field in the matter composition
significantly contributes in determining the fundamental structural
characteristics of any system. Several works related to the analysis
of the implications of electric charge on celestial formations in
GR, along with the alternative gravity theories can be seen in
literature
\cite{yousaf2021electrically,bhatti2023thin,bhatti2022analysis,yousaf2023cylindrical}.
Das et al. \cite{das2011isotropic} derived the solution to
Einstein-Maxwell field equations after the combination of static
internal structure and external Reissner-Nordstr\"{o}m spacetime.
Sunzu et al. \cite{sunzu2014charged} evaluated the structural
formation of strange stars and studied the impacts of electric
charge after the consideration of the relationship between mass and
radius. Murad \cite{murad2016some} established the charged
composition of matter for a strange star and studied the fundamental
features of the model. The relevance of electric field in
maintaining the stability of compact objects is also studied in
\cite{pant2011well,gupta2011class}. Further, The equation of state
(EoS) is important in analyzing the structural composition of
celestial formations. The anisotropic EoS is sufficient in
determining the solution for hydrostatic equilibrium equations.
Moreover, the polytropic EoS describes the connection between
pressure and density and yields the Lane-Emden equation as a
solution, which has significant relevance for characterizing the
anisotropic fluids \cite{heinzle2003dynamical,read2009constraints}.
Thirukkanesh and Ragel \cite{thirukkanesh2012exact} successfully
transformed Einstein field equations into two precise frameworks
within the context of polytropic EoS.

Herrera and Barreto \cite{herrera2013general} suggested the
generalized formation of the polytropic framework related to the
celestial objects comprising the anisotropic fluid. They analyzed
the effect of anisotropic pressure, energy density, and Tolman mass
on these structures. Thirukkanesh et al.
\cite{thirukkanesh2020model} derived a unique solution satisfying
the polytropic EoS. Their study demonstrated that the radial
component of pressure dominates the tangential pressure in parabolic
geometry. Ramos et al. \cite{ramos2021class} evaluated the Karmarkar
solutions through the consideration of anisotropic polytropes. They
solved the Lane-Emden for several sets of variables in isothermal
and non-isothermal compositions. They determined the specific
relations for Tolman mass and evaluated the effect of Karmarkar
constraint on the mass function. In analyzing the self-gravitating
compositions, fluid estimation is commonly considered along with the
premise of local isotropy. Nevertheless, higher densities of
self-gravitating fluids along with irregular primary stresses
anticipate the anisotropic essence of the fluid. The unusual periods
of transitions \cite{sokolov1980phase}, rotations
\cite{kippenhahn1990stellar}, and basic development of structure
\cite{herrera2009expansion,herrera2010cavity} may also cause the
emergence of anisotropy in the celestial objects. This suggests that
the self-gravitating composition of matter must contain two distinct
kinds of pressure components termed radial and tangential.
Consequently, the uneven components of pressure result in the
anisotropic matter distribution. Herrera and Santos brought this
finding to light in 1997 \cite{herrera1997local} which then
addressed by several researchers
\cite{maurya2019anisotropic,shamir2020stellar}.

General relativity, introduced by Albert Einstein, is widely
accepted being the most accurate interpretation of gravity.
Gravitational phenomena ranging from small scales to galactic
formations can be evaluated through this theory of gravity.
Subsequently, it has been acknowledged that GR fulfills the standards
suggested by solar system tests. The $\Lambda$-CDM framework, which was
developed within the framework of GR, is seen to be the most
pertinent framework for explaining the development of the universe.
Afterwards, several mathematical and experimental challenges
identified by scientific research prompted the generalization of GR.
The presence of dark matter at both stellar and galactic levels, in
addition to the emergence of singularity within a black hole and the
enigma related to the essence of dark energy are among the
unanswered questions in GR. Moreover, GR is thought to be
insufficient in elaborating the cosmic accelerated expansion.
Consequently, we can say that the modified form of GR may be helpful
in explaining the essence of these dark components. Significant
attempts have been made to provide a suitable gravitational
framework for the understanding of accelerating cosmic expansion.
There are primarily two ways that define this venture. The first
focuses on the configuration of matter that constitutes the major
portion of our universe. According to this, the cosmos is comprised
of an enigmatic negative pressure specifies as dark energy that
generates anti-gravitating pressure and ultimately contributes
significantly to the preservation and escalation of the cosmic
expansion. In this context, the gravitational equations are modified
through the addition of cosmological constant $\Lambda$.

However, after taking into account the particular dark energy
models, the accelerated cosmic expansion can also be described.
However, the cosmological constant problem arises in this case that
identifies the discrepancy between the measured and experimentally
calculated value of $\Lambda$. In another approach, the geometric
composition of spacetime associated with somewhat greater distances
is modified to find solutions for the accelerated development of the
universe. As a result, the idea of alternative gravity theories that
incorporate several different methods surfaced in the literature.
These theories are mainly related to the altered version of the
linear function specifying the curvature scalar
$R=g^{\alpha\beta}R_{\alpha\beta}$. The mathematical form of these
theories is mainly dependent on the extension of the gravitational
Lagrangian that entails a specific form $L_{GR}=R$ in the framework
of GR. After the consideration of the Gauss-Bonnet scalar $G$ and
the Ricci scalar $R$ as well as the effect caused by matter
dispersion that intercedes from the trace of energy-momentum tensor
as gravitational Lagrangian, establishes specific models relating to
these alternative theories.

The presence of dark enegry is also endorsed by the contemplation of
modified theories of gravity. In this regard, $f(R)$ theory is
typically thought of as the most straightforward modification in GR
\cite{buchdahl1970non,starobinsky2007disappearing}, in which the
generalized function of Ricci scalar is inserted in the place of
$L_{GR}=R$. The non-linear consequences of the curvature invariant
in the cosmic expansion have been reviewed after the consideration
of suitable $f(R)$ model \cite{sotiriou2010f,de2010f}. Amendola et
al. \cite{amendola2007conditions} determined specific limitations on
the viability of $f(R)$ dark energy frameworks and assessed their
endurance for the more precise depiction of cosmic expansion.
Capozziello et al. \cite{capozziello2015connecting} also discussed
inflation and dark energy eras in this theory. Nojiri and Odintsov
\cite{nojiri2006modified} proposed a $f(R)$ model for the
elaboration of dark energy that might be established after assuming
the FLRW spacetime. They assessed feasible $f(R)$ models for
distinguishing different phases of the universe. The addition of
particular relations between geometry and matter results in further
developed forms of $f(R)$ theory.

The work done by Bertolami et al. \cite{bertolami2007extra} has
significant relevance in developing the relationship between matter
and the composition of geometry in $f(R)$ theory. Their
approach suggested the combination between curvature invariant and
the matter Lagrangian in the form of $f(R, L_{m})$ gravity. As a
result of which, Harko et al. \cite{harko2011f} developed the
amazing alternative theory termed $f(R, \mathcal{T})$ gravity. The
consideration of Lagrangian functional as $f(R,\mathcal{T})$ is
regarded as a fascinating gravitational framework. This mathematical
framework employs a general mathematical formation to devise a
non-conserved distribution of matter yielding the production of an
additional force that pushes matter in non-geodesic paths
\cite{deng2015solar}. Alvarenga et al. \cite{alvarenga2013dynamics}
considered perturbations related to the FRW geometry within this
framework. Baffou et al. \cite{baffou2017late} studied late-time
cosmic evolution after incorporating the effects of Lagrange
multipliers in this theory. Bhatti et al.
\cite{bhatti2020stability,bhatti2023cylindrical,ur2024dynamically,bhatti2024construction,bhatti2022dynamical}
addressed the parameters determining the resilience of celestial
formations comprised of anisotropic fluid.

The incorporation of quantum effects in the formalism of $f(R,
\mathcal{T})$ theory may help us in explaining the phenomenon of
particle generation. This feature of $f(R, \mathcal{T})$ gravity is
important in astrophysical study as it explains the relationship
between alternative theory and quantum mechanics. The $f(R,
\mathcal{T})$ gravity is an appealing alternative to GR, that
provides numerous descriptions for cosmic events. However, the
challenges of developing the interpretation of a viable universe in
this theory can be seen in literature
\cite{velten2017cosmological,velten2021conserve}. It is determined
that this theory does not provide the appropriate explanation for
the cosmic expansion. In the context of these findings, Haghani and
Harko \cite{haghani2021generalizing} suggested the merged form of
two theories as $f(R, L_{m},\mathcal{T})$ gravity. This approach
resolves the shortcomings in preceding gravity theories leading to a
comprehensive description of complex cosmic phenomena. They
investigated the Newtonian limit of associated field equations and
proposed certain parameters for describing the accelerated cosmic
expansion after contemplating the low-velocity objects along with
weak fields of gravity. Zubair et al.
\cite{zubair2023thermodynamics} determined the cosmological
solutions incorporating de-Sitter and CDM frameworks, and
demonstrated their sustainability through fluctuations in $f(R,
L_{m}, \mathcal{T})$ theory. Naseer and his colleagues
\cite{naseer2024existence1,naseer2024existence} examined the
viability of different solutions representing cosmic objects in the
existence of Maxwell field. Some other interesting works on compact
fluid configurations within the framework of modified theories can
be found in
\cite{aj,ak,al,am,an,ao,ap,aq,ar,as,at,au,av,aw,ax,ay,ca,cb,cc,cd,ce,cf,cg,ch,ci,cj,ck,cl,cm,cn,co,cp,cq,cr,
cs,ct}.

Our manuscript is arranged in the following form: Section \textbf{2}
is related to the fundamental formalism of $f(R, L_{m},
\mathcal{T})$ gravity along with the essential terms linked with the
composition of fluid. The evolution equation and certain expressions
of different mass functions are also derived. The junction
conditions corresponding to $f(R, L_{m}, \mathcal{T})$ theory are
analyzed in the same section. We then present precise connections
between mass, conformal tensor, and Tolman mass. The significance of
all these components in interpreting the complexity is also
endorsed. The orthogonal decomposition of Riemann curvature tensor
yields the explanation for the fundamental characteristics of the
matter distribution in section \textbf{3}. A scalar quantity is
claimed as CF, which is of significant relevance in evaluating the
fundamental characteristics of complexity. Section \textbf{4}
addresses the vanishing complexity condition after considering two
certain frameworks proposed by Gokhroo and Mehra
\cite{gokhroo1994anisotropic} and polytropic EoS. Section \textbf{5}
discusses the implications of our findings in the context of stellar
structures. The main results are finally summarized in section
\textbf{6}.

\section{$f(R, L_{m}, \mathcal{T})$ formalism}

We establish the mathematical formalism of $f(R, L_{m},
\mathcal{T})$ gravity along with the consideration of spherically
symmetric anisotropic matter composition in the existence of
electromagnetic field. The modified equations of motion associated
with this extended gravity are solved in addition with the analysis
of some kinematical variables which have crucial significance in
determining the structure scalars. Furthermore, Darmois matching
constraints are assessed at the hypersurface. In the framework of
spherically charged anisotropic composition of matter, the
expressions for Misner-Sharp and Tolman masses are also established
that eventually lead to determining the relationship between
conformal tensor and fundamental parameters. The impact of
Misner-Sharp mass, electric field, and conformal tensor on the
anisotropic matter setup having non-homogeneous energy density
within the context of $f(R, L_{m}, \mathcal{T})$ theory is also
evaluated in this section.

\subsection{Derivation of modified field equations}

The action associated with the considered $f(R, L_{m}, \mathcal{T})$
gravity is obtained as \cite{naseer2024dynamic}
\begin{equation}\label{1}
S=\int \sqrt{-g}\left[\frac{f(R,L_{m},\mathcal{T})}{16
\pi}+L_E+L_{m}\right]d^{4}x,
\end{equation}
in which $L_{E}$ and $L_m$ are the Lagrangian densities of
electromagnetic field and usual fluid distribution, respectively,
and $g=|g_{\mu\nu}|$. The deviation of action given in \eqref{1} in
relation to $g_{\mu \nu}$ results in modified field equations as
described below
\begin{equation}\label{2}
\mathcal{G}_{\mu \nu}= 8 \pi \left( \mathcal{T}_{\mu \nu}^{(eff)}+E_{\mu\nu}\right),
\end{equation}
where the Einstein tensor is represented by $\mathcal{G}_{\mu \nu}$,
describing the structural arrangement of fluid, and the matter
contained in considered geometry is mentioned as $\mathcal{T}_{\mu
\nu}^{(eff)}$. Also, $E_{\mu\nu}$ specifies the energy-momentum
tensor associated with the electric field. The term
$\mathcal{T}_{\mu \nu}^{(eff)}$ is specifically classified as
follows
\begin{equation}\label{3}
\mathcal{T}_{\mu \nu}^{(eff)}=\frac{1}{f_{R}}\mathcal{T}_{\mu
\nu}^{(m)}+\mathcal{T}_{\mu \nu}^{(cr)},
\end{equation}
where $\mathcal{T}_{\mu \nu}^{(m)}$ specifies the matter
configuration and $\mathcal{T}_{\mu \nu}^{(cr)}$ represents
correction terms. Moreover, $\mathcal{T}_{\mu \nu}^{(cr)}$ is
described as
\begin{align}\nonumber
\mathcal{T}_{\mu \nu}^{(cr)}&=\frac{1}{8 \pi
f_{R}}\left[\frac{1}{2}\left(2f_{\mathcal{T}}+f
_{L_{m}}\right)\mathcal{T}_{\mu \nu}^{(m)}- \left(g_{\mu \nu} \Box
-\nabla_{\mu}\nabla_{\nu}\right)f_{R}
\right.\\\label{4}&\left.+\frac{1}{2}\left(f-Rf_{R}\right)g_{\mu
\nu}-\left(2f_{\mathcal{T}}+f _{L_{m}}\right)L_{m}g_{\mu \nu}+2
f_{\mathcal{T}}g^{\alpha \beta}\frac{\partial^{2}L_{m}}{\partial
g^{\mu \nu}\partial g^{\alpha \beta}}\right],
\end{align}
in which $f_{\mathcal{T}}=\frac{\partial f\left(R,
L_{m},\mathcal{T}\right)}{\partial \mathcal{T}}$,
$f_{L_{m}}=\frac{\partial f\left(R,
L_{m},\mathcal{T}\right)}{\partial L_{m}}$, and
$f_{R}=\frac{\partial f\left(R, L_{m},\mathcal{T}\right)}{\partial
R}$. Further, the D'Alembertian operator and covariant derivative
are mathematically stated as $\Box \equiv
\left(-g\right)^{-1/2}\partial_{\mu}\left(\sqrt{-g}g^{\mu
\nu}\partial_{\nu}\right)$ and $\nabla_{\mu}$, respectively. The
mathematical expression for $E_{\mu \nu}$ is
\begin{equation}\label{1c}
E_{\mu\nu}=\frac{1}{4\pi}\left(-S^{\alpha}_{\mu}S_{\nu\alpha}+\frac{1}{4}S^{\alpha\beta}S_{\alpha\beta}g_{\mu\nu}\right),
\end{equation}
in which $S_{\mu\nu}=\phi_{\mu; \nu}-\phi_{\nu; \mu}$ describes the
Maxwell field tensor satisfying the following equations
\begin{equation}\label{1ca}
S^{\mu\nu};_{\nu}=4\pi j^{\mu},\quad S_{[\mu\nu;\delta]}=0,
\end{equation}
with $\varphi_{\zeta}=\varphi(r)\delta^{0}_{\zeta}$ and $j^{\mu}$
indicating the four-potential and current density, respectively.

The energy-momentum tensor for self-gravitaing source is determined
by the inner composition of compact star in the framework of
$f(R,L_{m},\mathcal{T})$ gravity. The quantity $T^{\mu(m)}_{\nu}$
expressing the anisotropic fluid is specified as
\begin{equation}\label{5}
\mathcal{T}_{\nu}^{\mu(m)}=\rho u^{\mu} u_{\nu}-P
h_{\nu}^{\mu}+\Pi_{\nu}^{\mu},
\end{equation}
where
\begin{equation}\label{6}
\Pi_{\nu}^{\mu}=\Pi\left({l^{\mu}l_{\nu}} +\frac{1}{3}
h_{\nu}^{\mu}\right), \quad P=\frac{P_{r}+2 P_{\bot}}{3},
\end{equation}
\begin{equation}\label{7}
\Pi=P_{r}-P_{\bot}, \quad h_{\nu}^{\mu}=\delta_{\nu}^{\mu}-u^{\mu}
u_{\nu},
\end{equation}
and $\Pi$ indicates the anisotropic pressure. Also, $h_{\nu}^{\mu}$
is the projection tensor, and the principal pressure is divided into
radial ($P_{r}$) and tangential ($P_{\bot}$) components. Further,
$l^{\mu}$ being the four-vector, $\rho$ specifies the energy
density, and $u^{\mu}$ being the four-velocity.

The inner spherical geometry is defined in the subsequent form
\cite{bhatti2020gravastars}
\begin{equation}\label{8}
d s^{2}=e^{\xi(r)} d t^{2}-e^{\chi(r)} d
r^{2}-r^{2}\left(d^{2} \theta+\sin ^{2} \theta d^{2} \phi\right).
\end{equation}
Consequently, four-velocity and four-vector take the following form
\begin{equation}\label{9}
u^{\mu}=\left(e^{\xi/2}, 0,0,0\right), \quad l^{\mu}=\left(0,
e^{-\chi / 2}, 0,0\right),
\end{equation}
fulfilling
\begin{equation}\label{10}
l^{\mu}l_{\mu}=-1, \quad u^{\mu}u_{\mu}=1, \quad l^{\mu}u_{\mu}=0.
\end{equation}
Solving Eq. \eqref{1ca} (left) results in the following value of the
interior charge as
\begin{equation}\label{2c}
q(r)=\int_{0}^{r}\psi r^{2}e^{\frac{\chi}{2}}dr.
\end{equation}
in which $\psi$ represents the charge density.

The non-vanishing components of modified field equations associated
with $f(R,L_{m},\mathcal{T})$ gravity under the metric \eqref{8} are
described as
\begin{eqnarray}\label{9}
&&8\pi\left(\rho+\mathcal{T}_{0}^{0(cr)}-\frac{q^{2}}{8 \pi r^{4}}\right)=
\frac{1}{r^{2}}+\left(\frac{\chi^{\prime}}{2}-\frac{1}{r^{2}}\right)
e^{-\chi},
\\\label{10}
&&8\pi\left(-P_{r}+\mathcal{T}_{1}^{1(cr)}+\frac{q^{2}}{8 \pi
r^{4}}\right)=
\frac{1}{r^{2}}-\left(\frac{\xi^{\prime}}{r}+\frac{1}{r^{2}}\right)
e^{-\chi},
\\\label{11}
&&32
\pi\left(-P_{\bot}+\mathcal{T}_{2}^{2(cr)}-\frac{q^{2}}{8 \pi r^{4}}\right)=
e^{-\chi}\left\{\xi^{\prime}
\chi^{\prime}-2 \xi^{\prime \prime}-\xi^{\prime
2}+\frac{2\left(\chi^{\prime}-\xi^{\prime}\right)}{r}\right\},
\end{eqnarray}
in which $\mathcal{T}_{0}^{0(cr)}$, $\mathcal{T}_{1}^{1(cr)}$ and
$\mathcal{T}_{2}^{2(cr)}$ are the correction terms and their
mathematical expressions can be seen in Appendix. The hydrostatic
equilibrium condition is determined in the following way
\begin{align}\nonumber
&\frac{1}{2}\xi^{'}e^{-\chi}\left(\rho+P_{r}\right)+\frac{1}{2}\xi^{'}e^{\xi-\chi}\mathcal{T}^{00(cr)}-\frac{\xi^{'}e^{-\chi}q^{2}}{16
\pi
r^{4}}+\frac{1}{2}\xi^{'}\mathcal{T}^{11(cr)}+\frac{q^{2}\xi^{'}}{16
\pi r^{4}e^{\chi}}
+P_{r}^{'}e^{-\chi}+\mathcal{T}^{11'(cr)}\\\nonumber&+\left(\frac{q^{2}}{8
\pi
r^{4}e^{\chi}}\right)^{'}+\chi^{'}\mathcal{T}^{11(cr)}+\frac{q^{2}\chi^{'}}{8
\pi r^{4}e^{\chi}}-\frac{q^{2}r}{8 \pi
r^{6}e^{\chi}}-\frac{e^{-\chi}}{r}P_{\bot}-re^{-\chi}\mathcal{T}^{22(cr)}+\frac{1}{r}P_{r}e^{-\chi}\\\label{12}&
+\frac{1}{r}\mathcal{T}^{11(cr)}+\frac{q^{2}}{4 \pi
r^{5}e^{\chi}}=Z^{\star},
\end{align}
where the value of $Z^{\star}$ is mentioned in Appendix. This is
named as TOV equation, characterizing the evolution of anisotropic
fluid configuration is important in studying the structural changes
of the system under consideration. Equation \eqref{10} yields the
following value
\begin{equation}\label{13}
\xi^{'}=\frac{8 \pi r^{3}\left(P_{r}-\mathcal{T}^{1(cr)}_{1}\right)+\left(m-\frac{q^{2}}{4}\right)}{r(r-2m+\frac{q^{2}}{2})}.
\end{equation}
The insertion of $\xi'$ from above equation implies the following
form of non-conservation equation
\begin{align}\nonumber
P_{r}^{'}&=-\left(\frac{8 \pi r^{3}\left(P_{r}-\mathcal{T}^{1(cr)}_{1}\right)+\left(m-\frac{q^{2}}{4}\right)}{r(r-2m+\frac{q^{2}}{2})}\right)\left(\frac{-1}{2}\left(\rho+P_{r}\right)
-\frac{e^{\xi}}{2}\mathcal{T}^{00(cr)}-\frac{q^{2}}{8 \pi r^{4}}\right)
\\\nonumber& -e^{\chi} \mathcal{T}^{11'(cr)}-e^{\chi}\left(\frac{q^{2}}{8 \pi r^{4}e^{\chi}}\right)^{'}-\chi^{'}e^{\chi}\mathcal{T}^{11(cr)}-\frac{q^{2}\chi^{'}}{8 \pi r^{4}}+\frac{q^{2}r}{8 \pi r^{6}}+r\mathcal{T}^{22(cr)}-\frac{2}{r}e^{\chi}\\\label{14}& \times \mathcal{T}^{11(cr)}+\frac{P_{\bot}}{4}-\frac{2P_{r}}{r}-\frac{3 q^{2}}{4 \pi r^{5}}+e^{\chi}Z^{\star}.
\end{align}

The hypersurface having three dimensions accurately aligns the
internal and external regions of compact structures. The matching
restrictions proposed by Darmois \cite{darmois1927memorial} are
regarded as the appropriate depiction for smooth matching of
distinct sectors. The Reissner-Nordstr\"{o}m metric is considered
for the external sector as
\begin{equation}\label{22}
ds^{2}=\left(1-\frac{2M}{r}+\frac{Q^{2}}{r^{2}}\right)dt^{2}-\frac{dr^{2}}{\left(1-\frac{2M}{r}
+\frac{Q^{2}}{r^{2}}\right)}
-r^{2}d\theta^{2}-r^{2}\sin^{2}\theta d\phi^{2},
\end{equation}
where $M$ is the mass related to the external geometry and $Q$
specifies the charge enclosed by this. Moreover, the fundamental
forms of matching conditions are mentioned as follow\\
$\bullet$ The spacetimes relating to internal and external sectors
should be continuous at the hypersurface described as
\begin{equation}\label{D1}
[ds^{2}_{+}]_{\Sigma}=[ds^{2}]_{\Sigma}=[ds^{2}_{-}]_{\Sigma}.
\end{equation}
$\bullet$ The extrinsic curvature ought to be continuous at $\Sigma$
as
\begin{equation}\label{D2}
[K_{ef}]_{\Sigma}=[K_{ef}]_{+}=[K_{ef}]_{-},
\end{equation}
The mathematical representation of external curvature is as follows
\begin{equation}\label{D3}
K^{\pm}_{ef}=-n^{\pm}_{\alpha}\left[\frac{\partial^{2}x_{\alpha}}{\partial\zeta^{e}\partial\zeta^{f}}
+\Gamma^{\alpha}_{\vartheta\lambda}\frac{\partial
x^{\vartheta}}{\partial \zeta^{e}}\frac{\partial x^{\lambda}}{\partial
\zeta^{f}}\right]_{\Sigma}.
\end{equation}
It is crucial to note that $\zeta^{e}$ represents the interior
coordinates and $n^{\pm}_{\alpha}$ specifies the normal vector associated
with the horizon. The contemplation of Eqs. \eqref{D1}-\eqref{D3} at
$r=r_{\Sigma}$ results
\begin{equation}\label{D4}
e^{\xi_{\Sigma}}=1-\frac{2M}{r_{\Sigma}}+\frac{Q^{2}}{r^{2}_{\Sigma}},
\quad
e^{-\chi_{\Sigma}}=1-\frac{2M}{r_{\Sigma}}+\frac{Q^{2}}{r^{2}_{\Sigma}},
\quad [P_{r}]_{\Sigma}=\mathcal{F}_{0},
\end{equation}
where the expression for $\mathcal{F}_{0}$ is provided in Appendix.

\subsection{Relationship between curvature tensor and certain physical quantities}

The association between different quantities is expressed as follows
\begin{equation}\label{7c}
R_{\delta e  \mu}^{\lambda}=\mathcal{C}_{\delta e
\mu}^{\lambda}+\frac{1}{2} R_{e}^{\lambda} g_{\delta
\mu}-\frac{1}{2} R_{\delta e} \delta_{\mu}^{\lambda}+
\frac{1}{2} R_{\delta \mu} \delta_{e}^{\lambda} -\frac{1}{2}
R_{\mu}^{\lambda} g_{\delta e}-\frac{1}{6}
R\left(\delta_{\mu}^{\lambda} g_{\delta e}-g_{\delta \mu}
\delta_{e}^{\lambda}\right).
\end{equation}
The conformal tensor is regarded as unique
factor of curvature tensor which claims the description of
propagating waves without the presence of any substance. The most
distinguishing feature of conformal tensor is the trace-free nature.
The Riemann tensor consists of both an electric and magnetic
factors. The independence of line expansion in fluid flow of spherically symmetric distribution of substance
implies the ignorance of magnetic impacts. Eventually, the collapse
spins along the fluid distribution. The electric component related
to the conformal tensor is derived after the consideration of the
expression given below
\begin{align}\label{13}
\mathcal{E}_{\delta \alpha}&=\mathcal{C}_{\delta\beta\alpha\tau}
u^{\beta} u^{\tau}, \quad \beta, \tau=0,1,2,3,\\\label{13}
\mathcal{C}_{\delta\beta\alpha\tau}&=\left(g_{\delta\beta
e\varrho}g_{\alpha\tau\sigma\lambda} -\eta_{\delta\beta e\sigma}
\eta_{\alpha\tau\sigma\lambda}\right)u^{e}u^{\varrho}
\mathcal{E}^{\sigma}\mathcal{E}^{\lambda}.
\end{align}
The alternative form is
\begin{equation}\label{13}
\mathcal{E}_{\delta \alpha}=\in\left(l_{\delta} l_{\alpha}+\frac{1}{3}
h_{\delta \alpha}\right),
\end{equation}
where $\in$ is the Weyl scalar having the value as
\begin{equation}\label{1a}
\in=\frac{-e^{-\chi}}{4}\left[\xi^{\prime
\prime}+\frac{\xi^{\prime2}-\chi^{\prime}
\xi^{\prime}}{2}-\frac{\xi^{\prime}-\chi^{\prime}}{r}
+\frac{2\left(1-e^{\chi}\right)}{r^{2}}\right],
\end{equation}
with the restraints defined as
\begin{equation}\label{13}
\in_{\xi}^{\xi}=0, \quad \in_{\xi \lambda}=\in_{(\xi \lambda)}, \quad
\in_{\xi \alpha} u^{\alpha}=0.
\end{equation}

\subsection{Two different mass functions}

The distribution of matter is analyzed through the consideration of
relations for mass proposed by Misner-Sharp
\cite{misner1964relativistic} and Richard C. Tolman
\cite{tolman1930use}. Although, these mathematical models predict
similar results at the horizon, yet they claim different
explanations of energy related to inner sector due to the
distribution of matter. The fundamental effects of gravitational
collapse is determined through the assumption of Misner-Sharp mass
\cite{wilson1971numerical,bruenn1985stellar}, while, the expression
for mass recommended by Tolman \cite{bonnor2001interactions}
considered as appropriate gravitating mass. The former mass
associated with spherically symmetric system in the existence of
electric field is calculated as
\begin{equation}\label{15}
m=\frac{r}{2}\left(1-e^{-\chi}\right)+\frac{q^{2}}{2 r}.
\end{equation}
Differentiating the above equation along with the contemplation of
Eq. \eqref{9} concludes
\begin{equation}\label{16}
m=4 \pi \int_{0}^{r} r^{2}\left(\rho+\mathcal{T}_{0}^{0(cr)}-\frac{q^{2}}{8 \pi r^{4}}\right) d r + \int_{0}^{r} \frac{q q^{'}}{r} dr - \int_{0}^{r} \frac{q^{2}}{2 r^{2}}dr.
\end{equation}
The consideration of Eqs. (\ref{9})-(\ref{11}) , along with the
expression for Misner-Sharp mass results
\begin{align}\label{17}
m&=\frac{4 \pi r^{3}}{3}\left[\rho-2
P_{\bot}-P_{r}+\mathcal{T}_{0}^{0(cr)}+2\mathcal{T}_{2}^{2(cr)}+\mathcal{T}_{1}^{1(cr)}\right]-\frac{4
q^{2}}{3r}+\frac{q^{2}}{4}+\frac{r^{3}}{3} \in.
\end{align}

A new form of the earlier mentioned mass function can be obtained in
the context of the value of $\in$ given in Eq. \eqref{1a} by
\begin{align}\nonumber
m&=\frac{4 \pi r^{3}}{3}\left[\rho-2
P_{\bot}-P_{r}+\mathcal{T}_{0}^{0(cr)}+2\mathcal{T}_{2}^{2(cr)}+\mathcal{T}_{1}^{1(cr)}\right]-\frac{4
q^{2}}{3r}+\frac{q^{2}}{4}\\\label{17b}&-\frac{ r^{3}
e^{-\chi}}{12}\left[\xi^{\prime
\prime}+\frac{\xi^{\prime2}-\chi^{\prime}
\xi^{\prime}}{2}-\frac{\xi^{\prime}-\chi^{\prime}}{r}
+\frac{2\left(1-e^{\chi}\right)}{r^{2}}\right].
\end{align}
The comparison between Eqs. \eqref{16} and \eqref{17} results in a
novel representation of $\in$, given below
\begin{align}\nonumber
\in&= -\frac{4 \pi}{r^{3}}\int_{0}^{r}
r^{3}\left(\rho+\mathcal{T}_{0}^{0(cr)}\right)^{\prime} d r+4
\pi\left(P_{r}+2P_{\bot}-\mathcal{T}_{1}^{1(cr)}-2
\mathcal{T}_{2}^{2(cr)}\right)\\\label{18}&+\frac{3 q^{2}}{4
r^{3}}-\frac{4 q^{2}}{r^{4}}-\frac{6}{r^{3}}\int
\frac{q^{2}}{r^{2}}dr+\frac{3}{r^{3}}\int \frac{q q^{'}}{r}dr.
\end{align}
The relationship between conformal tensor and distinct
characteristics including non-homogenous energy density and
anisotropic pressure of spherically symmetric configuration in the
presence of electric charge is described by this equation.
Eventually, Eq. \eqref{17} takes the following form after the
insertion of the value of $\in$ by
\begin{equation}\label{19}
m=\frac{4 \pi
r^{3}}{3}\left[\rho+\mathcal{T}_{0}^{0(cr)}-\mathcal{T}_{2}^{2(cr)}\right]-\frac{4
\pi}{3} \int_{0}^{r}
r^{3}\left(\rho+\mathcal{T}_{0}^{0(cr)}\right)^{\prime} d r-\frac{8 q^{2}}{3 r}-2\int \frac{q^{2}}{r^{2}}dr +\int \frac{q q^{'}}{r}dr.
\end{equation}

The existence of non-uniform energy density, electric charge and
anisotropic distribution of pressure imply substantial effect on fluid dispersion
via Weyl tensor in $f(R, L_{m},\mathcal{T})$ theory. While after the
consideration of uniform mass dispersion, non-uniformity is used to
establish the desired variation which is signified through the mass
relation given in Eq. \eqref{19}. Tolman \cite{tolman1930use}
suggested a significant description of formation of energy
related to the matter. The Tolman mass related to the spherically symmetric
matter composition is written as
\begin{equation}\label{20}
m_{T}=4 \pi \int_{0}^{r_{\Sigma}} r^{2} e^{(\xi+\chi) / 2}\left[ \rho+P_{r}+2
P_{\bot}\right] d r.
\end{equation}
Bhatti et al. \cite{bhatti2020stability,bhatti2019tolman} derived
the expression for mass function related to the spherical system in
the context of a modified theory.  The derivation of mass related to
spherical composition of matter in the existence of electric field
is now illustrated in $f(R, L_{m}, \mathcal{T})$ theory as
\begin{equation}\label{21}
m_{T}=4 \pi \int_{0}^{r} r^{2} e^{(\xi+\chi) / 2}\left[ \rho+P_{r}+2
P_{\bot}\right] d r,
\end{equation}
evaluating the behaviour of inertial mass denoted as $m_{T}$,
explained in \cite{herrera2009structure,herrera1997local}. The
substitution of Eqs. (\ref{9})-(\ref{11}) into \eqref{21} results in
\begin{equation}\label{22}
m_{T}=e^{\frac{\xi-\chi}{2}} \frac{\xi^{\prime}
r^{2}}{2}+4 \pi \int_{0}^{r} r^{2}
e^\frac{\xi+\chi}{2}\left(\mathcal{T}_{1}^{1(cr)}-\mathcal{T}_{0}^{0(cr)}+2\mathcal{T}_{2}^{2(cr)}-\frac{q^{2}}{2 \pi r^{4}}\right) d r.
\end{equation}
In the context of the value of $\xi^{\prime}$, the above equation
takes the following form
\begin{align}\nonumber
m_{T}&=r e^{\frac{\xi-\chi}{2}}\left[\frac{4 \pi
r^{3}\left(P_{r}-\mathcal{T}_{1}^{1(cr)}\right)+\left(m-\frac{q^{2}}{r}\right)}
{r-2 m +\frac{q^{2}}{r}}\right]\\\label{23} &+4 \pi \int_{0}^{r}
r^{2}
e^{\frac{\xi+\chi}{2}}\left(\mathcal{T}_{1}^{1(cr)}-\mathcal{T}_{0}^{0(cr)}+2\mathcal{T}_{2}^{2(cr)}-\frac{q^{2}}{2
\pi r^{4}}\right)dr,
\end{align}
emphasizing the significance of $m_{T}$, also termed as an inertial
mass. Consequentially, Tolman mass is expressed as
\begin{align}\nonumber
m_{T}&=M\left(\frac{r}{r_{\Sigma}}\right)^{3}+\frac{r^{3}}{2}\int^{r}_{0}
\frac{e^{\frac{\xi -
\chi}{2}}}{r}\left(\xi^{''}+\frac{\xi^{'}}{r}+\frac{\xi^{'2}}{2}-\frac{\xi^{'}\chi^{'}}{2}\right)dr
+ 4 \pi \int ^{r}_{0}
r^{2}e^\frac{{\xi+\chi}}{2}\\\nonumber&\times\left(\mathcal{T}_{1}^{1(cr)}-\mathcal{T}_{0}^{0(cr)}+2\mathcal{T}_{2}^{2(cr)}
-\frac{q^{2}}{2 \pi r^{4}}\right)dr+4r^{3} \int^{r}_{0}
\frac{q^{2}e^{\frac{\xi +
\chi}{2}}}{r^{4}(r+\frac{q^{2}}{r})}dr+r^{3}
\\\nonumber& \times  \int^{r}_{0}\frac{1}{r^{3}(r+\frac{q^{2}}{r})} \left[4 \pi
r^{3}e^{\frac{\xi+\chi}{2}}\left(2\mathcal{T}_{2}^{2(cr)}+\frac{3
q^{2}}{8 \pi r^{4}} \right)+\frac{r}{2}e^\frac{{\xi
+\chi}}{2}-\frac{r^{3}}{2}e^{\frac{\xi-\chi}{2}}\right.\\\label{24}&\left.\times
\left(\frac{2
\chi^{'}}{r}+\frac{1}{r^{2}}+\frac{\xi^{'}}{r}+\frac{\xi^{'}\chi^{'}}{2}-\xi^{''}-\frac{\xi^{'2}}{2}\right)\right]dr
-r^{3}\int^{r}_{0}\frac{e^{\frac{\xi+\chi}{2}}}{r+\frac{q^{2}}{r}}\in
dr.
\end{align}
The preceding equation turns to the following form after the
consideration of Eq. \eqref{18} as
\begin{align}\nonumber
m_{T}&=M\left(\frac{r}{r_{\Sigma}}\right)^{3}+\frac{r^{3}}{2}\int^{r}_{0}
\frac{e^{\frac{\xi -
\chi}{2}}}{r}\left(\xi^{''}+\frac{\xi^{'}}{r}+\frac{\xi^{'2}}{2}-\frac{\xi^{'}\chi^{'}}{2}\right)dr
+ 4 \pi \int ^{r}_{0}
r^{2}e^\frac{{\xi+\chi}}{2}\\\nonumber&\times\left(\mathcal{T}_{1}^{1(cr)}-\mathcal{T}_{0}^{0(cr)}
+2\mathcal{T}_{2}^{2(cr)} -\frac{q^{2}}{2 \pi r^{4}}\right)dr+4r^{3}
\int^{r}_{0} \frac{q^{2}e^{\frac{\xi +
\chi}{2}}}{r^{4}(r+\frac{q^{2}}{r})}dr+r^{3}
\\\nonumber& \times \int^{r}_{0} \frac{1}{r^{3} (r+\frac{q^{2}}{r})} \left[4 \pi
r^{3}e^{\frac{\xi+\chi}{2}}\left(2\mathcal{T}_{2}^{2(cr)}+ \frac{3
q^{2}}{8 \pi r^{4}} \right)+\frac{r}{2}e^\frac{{\xi
+\chi}}{2}-\frac{r^{3}}{2}e^{\frac{\xi-\chi} {2}}\left(\frac{2
\chi^{'}}{r}+\frac{1}{r^{2}}\right.\right.\\\nonumber&\left.\left.+\frac{\xi^{'}}{r}+\frac{\xi^{'}\chi^{'}}
{2}-\xi^{''}-\frac{\xi^{'2}}{2}\right)\right]dr
-r^{3}\int^{r}_{0}\frac{e^{\frac{\xi+\chi}{2}}}{r+\frac{q^{2}}{r}}
\left[-\frac{4 \pi}{r^{3}}\int_{0}^{r}
r^{3}\left(\rho+\mathcal{T}_{0}^{0(cr)}\right)^{\prime} d
r\right.\\\label{25}&\left.+4
\pi\left(P_{r}+2P_{\bot}-\mathcal{T}_{1}^{1(cr)}-2
\mathcal{T}_{2}^{2(cr)}\right)+\frac{q^{2}}{2r^{3}}-\frac{6}{r^{3}}\int
\frac{q^{2}}{r^{2}}dr+\frac{3}{r^{3}}\int \frac{q q^{'}}{r}dr\right]
dr.
\end{align}
This equation is of significant relevance in evaluating the effect
of correction terms, anisotropic distribution of pressure and non-uniform energy
density on Tolman mass. As a result, it can be asserted that
this mathematical expression explains the consequences of non-uniform energy
density in addition to the anisotropic essence of pressure on Tolman
mass in $f(R, L_{m}, \mathcal{T})$ theory .

\section{Orthogonal splitting of the Riemann tensor}
Particular fundamental variables are considered for the elaboration
of characteristics associated with the fluid distribution. These
parameters are obtained by the orthogonal decomposition of Riemann
curvature tensor. Herrera \cite{herrera2018new} studied the basic
characteristics of anisotropic matter distribution after
contemplating the structural scalars. These are the trace and
trace-free components of particular tensors having significant relevance in
analyzing the composition of the system. In the current framework, structural
scalars are employed to assess the CF corresponding to the spherical
matter. We commence with the subsequent tensors
\cite{herrera2009structure,herrera2011role} given by
\begin{align}
Y_{e \omega}&=R_{\omega\delta e \alpha}u^{\delta}u^{\alpha},
\\\label{7a}
Z_{e \omega}&={}_{\star}R_{\omega\delta e \alpha} u^{\delta}
u^{\alpha}=\frac{1}{2} \eta_{\omega\delta\lambda\alpha} R_{e
\beta}^{\delta\alpha} u^{\lambda} u^{\beta},
\\\label{7b}
X_{e \omega }&= R{}^{\star}_{\omega\delta e \alpha}
u^{\delta} u^{\alpha}=\frac{1}{2} \eta_{\omega\delta}^{\lambda\alpha}
R_{\lambda\alpha e \beta}^{*} u^{\delta} u^{\beta} \text {, }
\end{align}
where $\star$ enumerates the dual tensor and
$\eta_{\omega\delta}^{\lambda\alpha}$ represents the Levi-Civita
symbol whose different values specify distinct variation in addition with
$R_{\zeta\lambda\delta\mu}^{\star}=\frac{1}{2}
\eta_{\vartheta\alpha\delta\mu} R_{\zeta\lambda}^{\vartheta\alpha}$.
The orthogonal splitting of the Riemann curvature tensor is
represented through the consideration of these tensors
\cite{gomez2008kerr}. The review of Eq. \eqref{7c}, along with
modified field equations concludes
\begin{equation}
R_{e \mu}^{\lambda \zeta}=C_{e \mu}^{\lambda \zeta}+28 \pi
\mathcal{T}^{[\lambda}_{[e}\delta^{\zeta]}_{\mu]}+8 \pi
\mathcal{T}^{(eff)}\left(\frac{1}{3}\delta^{\lambda}_{[e}\delta^{\zeta}_{\mu]}-\delta^{[\lambda}_{[e}\delta^{\zeta]}_{\mu]}
\right).
\end{equation}
The substitution of Eq. \eqref{2} in above equation results in the
subsequent expression for curvature tensors as
\begin{equation}
R_{e \mu}^{\lambda \zeta}=R_{(I) e \mu}^{\lambda
\zeta}+R_{(II) e \mu}^{\lambda \zeta}+R_{(III) e
\mu}^{\lambda \zeta}+R_{(IV) e \mu}^{\lambda
\zeta}+R_{(V) e \mu}^{\lambda \zeta},
\end{equation}
where
\begin{align}\label{S1}
R_{(I) e \mu}^{\lambda \zeta}&=16 \pi \rho u^{[\lambda} u_{[e}
\delta^{\zeta]}_{\mu]}- 16 P
h^{[\lambda}_{[e}\delta^{\zeta]}_{\mu]}+ 8\pi
\left(\rho-3P\right)\left(\frac{1}{3}\delta^{\lambda}_{[e}\delta^{\zeta}_{\mu]}-\delta^{[\lambda}_
{[e}\delta^{\zeta]}_{\mu]} \right),
\\\label{S2}
R_{(II) e \mu}^{\lambda
\zeta}&=16\Pi^{[\lambda}_{[e}\delta^{\zeta]}_{\mu]},
\\\label{S3}
R_{(III) e \mu}^{\lambda \zeta}&=4 u^{[\lambda} u_{[e}
\mathcal{E}^{\zeta]}_{\mu]}-\eta^{\lambda \zeta}_{\beta}\eta_{\alpha
e \mu}\mathcal{E}^{\beta \alpha},
\\\nonumber
R_{(IV) e \mu}^{\lambda \zeta}&=\frac{\delta^{\zeta}_{\mu}}{16 \pi
f_{R}}\left[f_{\mathcal{T}}\mathcal{T}^{\lambda
(m)}_{e}+\frac{1}{2}f_{L_{m}}\mathcal{T}^{\lambda
(m)}_{e}-\delta^{\lambda}_{e}\Box f_{R}+\nabla
^{\lambda}\nabla_{e}f_{R}+\frac{1}{2}\delta^{\lambda}_{e}f\right.\\\nonumber&\left.-\frac{1}{2}\delta^{\lambda}_{e}
R
f_{R}-2\delta^{\lambda}_{e}L_{m}f_{\mathcal{T}}-\delta^{\lambda}_{e}L_{m}f_{L_{m}}+2
f_{\mathcal{T}}g^{\alpha\beta}\frac{\partial ^{2}L_{m}}{\partial
\delta^{\lambda}_{e}\partial g^{\alpha
\beta}}\right]-\frac{\delta^{\zeta}_{e}}{16 \pi
f_{R}}\left[\mathcal{T}^{\lambda(m)}_{\mu}\right.\\\nonumber&\left.
\times
f_{\mathcal{T}}+\frac{1}{2}f_{L_{m}}\mathcal{T}^{\lambda(m)}_{\mu}-\delta^{\lambda}_{\mu}\Box
f_{R}+\nabla ^{\lambda}\nabla
_{\mu}f_{R}+\frac{1}{2}\delta^{\lambda}_{\mu}f-\frac{1}{2}\delta^{\lambda}_{\mu}R
f_{R}-\delta^{\lambda}_{\mu}L_{m}f_{\mathcal{T}}\right.\\\label{S4}&\left.-\delta^{\lambda}_{\mu}L_{m}
f_{L_{m}}+2f_{\mathcal{T}}\frac{g^{\alpha\beta}\partial^{2}L_{m}}{\partial
\delta^{\lambda}_{\mu}\partial g^{\alpha
\beta}}\right]+\frac{1}{2}E^{\lambda}_{e}\delta^{\zeta}_{\mu}-\frac{1}{2}E^{\lambda}_{\mu}\delta^{\zeta}_{e},
\\\nonumber
R_{(V) e \mu}^{\lambda \zeta}&=\frac{\delta^{\lambda}_{e}}{16 \pi
f_{R}}\left[f_{\mathcal{T}}\mathcal{T}^{\zeta(m)}_{\mu}+\frac{1}{2}f_{L_{m}}\mathcal{T}^{\zeta(m)}_{\mu}
-\delta^{\zeta}_{\mu}\Box f_{R}+\nabla ^{\zeta}\nabla
_{\mu}f_{R}+\frac{1}{2}\delta^{\zeta}_{\mu}f\right.\\\nonumber&\left.-\frac{1}{2}\delta^{\zeta}_{\mu}R
f_{R}-2\delta^{\zeta}_{\mu}L_{m}f_{\mathcal{T}}-\delta^{\zeta}_{\mu}L_{m}f_{L_{m}}
+2\frac{g^{\alpha \beta}\partial^{2}L_{m}}{\partial \delta
^{\zeta}_{\mu}\partial g^{\alpha \beta}}f_{\mathcal{T}}\right]
-\frac{\delta^{\lambda}_{\mu}}{16 \pi
f_{R}}\left[\mathcal{T}^{\zeta(m)}_{e}\right.\\\nonumber&\left.\times
f_{\mathcal{T}}+\frac{f_{L_{m}}}{2}\mathcal{T}^{\zeta(m)}_{e}-\delta^{\zeta}_{e}\Box
f_{R}+\nabla ^{\zeta}\nabla
_{e}f_{R}+\frac{1}{2}\delta^{\zeta}_{e}f-\frac{1}{2}\delta^{\zeta}_{e}R
f_{R}-2\delta^{\zeta}_{e}L_{m}\right.\\\label{S5}&\left. \times
f_{\mathcal{T}}-\delta^{\zeta}_{e}L_{m}f_{L_{m}}+\frac{2 g^{\alpha
\beta}\partial^{2}L_{m}}{\partial \delta^{\zeta}_{e}\partial
g^{\alpha
\beta}}f_{\mathcal{T}}\right]+\frac{1}{2}E^{\zeta}_{\mu}\delta^{\lambda}_{e}-\frac{1}{2}E^{\zeta}_{e}\delta^{\lambda}_{\mu}.
\end{align}
Several other properties are specified in the subsequent form
\begin{equation}
\epsilon_{\mu \zeta \beta}=v^{\alpha}\eta_{\alpha\mu\zeta\beta},
\quad \epsilon_{\mu \zeta \nu}v^{\nu}=0,
\end{equation}
and
\begin{equation}
\epsilon^{\mu \zeta \tau} \epsilon_{\tau \beta\varphi}=\delta_{\beta}^{\zeta} h_{\varphi}^{\mu}-\delta_{\beta}^{\mu} h_{\varphi}^{\zeta}+u_{\beta}\left(u^{\mu} \delta_{\varphi}^{\zeta}-u^{\zeta} \delta_{\varphi}^{\mu}\right).
\end{equation}
The electric component of conformal tensor plays an important role
in orthogonal decomposition of curvature tensor associated with the
charged spherical composition of matter. These certain tensors $X_{e
\omega}$, $Y_{e \omega }$, and $Z_{e \omega}$, are derived in the
following form
\begin{align}\\\label{53}
X_{e \omega}&=4 \pi \Pi_{e \omega}-E_{e \omega}-\frac{q^{2}}{2r^{4}}\left(\delta_{e}\delta_{\omega}+\frac{1}{3}h_{e \omega}\right)+\frac{8 \pi}{3} \rho h_{e \omega}+\mathcal{N}_{e \omega}^{(A)},
\\\label{54}
Y_{e \omega}&=4 \pi \Pi_{ e \omega}+E_{e \omega}+\frac{q^{2}}{r^{4}}h_{e \omega}-\frac{q^{2}}{2 r^{4}}\left(\delta_{e \omega}+\frac{1}{3}h_{e \omega}\right)+\frac{4 \pi}{3}\left(\rho+3P\right)h_{e \omega}+\mathcal{N}_{e \omega}^{(B)},
\\\label{55}
Z_{e \omega}&=\mathcal{N}_{e \omega}^{(C)}.
\end{align}
The values of $\mathcal{N}_{e \omega}^{(A)}$, $\mathcal{N}_{e
\omega}^{(B)}$ and $\mathcal{N}_{e \omega}^{(C)}$ are written in
Appendix. The tensors mentioned in these equations can be helpful in
the thorough study of a system's evolution
\cite{herrera2009structure}. Moreover, the subsequent form of trace
and trace-free parts of these scalars are used to evaluate the key
features of spherical configuration of charged matter
\begin{equation}\label{55b}
X_{T}=8 \pi \rho+\frac{q^{2}}{2r^{4}}+F,
\end{equation}
and the expression for $F$ can be seen in Appendix. Also, the related
trace-free component is
\begin{equation}\label{56}
X_{TF}=\frac{8 \pi \Pi}{3}-\frac{q^{2}}{2r^{4}}-\frac{2}{3}\in.
\end{equation}
The insertion of expression for $\in$ in the above equation results
\begin{align}\nonumber
X_{TF}&=\frac{8 \pi \Pi}{3}-\frac{q^{2}}{2r^{4}}+\frac{8 \pi}{3
r^{3}}\int_{0}^{r}
r^{3}\left(\rho+\mathcal{T}_{0}^{0(cr)}\right)^{\prime} d r-\frac{8
\pi}{3}\left(P_{r}+2P_{\bot}-\mathcal{T}_{1}^{1(cr)}\right.\\\label{57}&\left.-2
\mathcal{T}_{2}^{2(cr)}\right)-\frac{ q^{2}}{2 r^{3}}+\frac{8
q^{2}}{3 r^{4}}+\frac{4}{r^{3}}\int
\frac{q^{2}}{r^{2}}dr-\frac{2}{r^{3}}\int \frac{q q^{'}}{r}dr.
\end{align}

Furthermore,
\begin{equation}\label{58}
Y_{T}=4\pi \rho+\frac{3 q^{2}}{r^{4}}+D,
\end{equation}
where the vale of $D$ can be seen in Appendix. The analogous
trace-free component is
\begin{equation}\label{59}
Y_{TF}=\frac{8 \pi}{3}\Pi+\frac{2}{3}\in-\frac{q^{2}}{2 r^{4}}+L_{e \omega},
\end{equation}
where $L_{e \omega}=\frac{N_{e \omega}}{\delta_{e}
\delta_{\omega}+\frac{1}{3} h_{e \omega}}$. After the contemplation
of the value of $\in$ given in Eq. \eqref{18} concludes
\begin{align}\nonumber
Y_{TF}&=\frac{8 \pi}{3}\Pi-\frac{q^{2}}{2 r^{4}}+L_{e \omega}-\frac{8 \pi}{3 r^{3}}\int_{0}^{r}
r^{3}\left(\rho+\mathcal{T}_{0}^{0(cr)}\right)^{\prime} d r+\frac{8 \pi}{3}\left(P_{r}+2P_{\bot}-\mathcal{T}_{1}^{1(cr)}\right.\\\label{60}&\left.-2
\mathcal{T}_{2}^{2(cr)}\right)+\frac{ q^{2}}{2 r^{3}}-\frac{8 q^{2}}{3 r^{4}}-\frac{4}{r^{3}}\int \frac{q^{2}}{r^{2}}dr+\frac{2}{r^{3}}\int \frac{q q^{'}}{r}dr.
\end{align}
The anisotropic essence of pressure is described in subsequent form
through the trace-free parts given in Eqs. (\ref{57}) and (\ref{60})
as
\begin{equation}\label{61}
X_{TF}+Y_{TF}=\frac{16 \pi \Pi}{3}-\frac{2 q^{2}}{r^{4}}+L_{e \omega}.
\end{equation}
The consideration of Eq. \eqref{60} in \eqref{25} results the subsequent
relation between $Y_{TF}$ and the Tolman mass as
\begin{align}\nonumber
m_{T}&=M\left(\frac{r}{r_{\Sigma}}\right)^{3}+r^{3}
\int_{r}^{r_{\Sigma}}
\frac{e^{\frac{\xi+\chi}{2}}}{\left(r+\frac{q^{2}}{r}\right)}\left(Y_{T
F}-L_{e \omega}+2\mathcal{T}_{2}^{2(cr)}+\frac{q^{2}}{8 \pi
r^{4}}\right)dr+4 r^{3}\\\nonumber& \times \int_{r}^{r_{\Sigma}}
\frac{q^{2}e^{\frac{\xi+\chi}{2}}}{r^{4}\left(r+\frac{q^{2}}{r}\right)}dr
+4 \pi \int_{r}^{r_{\Sigma}} r^{2} e^{\frac{\xi+\chi}{2}}\left(
2\mathcal{T}_{2}^{2(D)}+\mathcal{T}_{1}^{1(D)}-\mathcal{T}_{0}^{0(D)}-\frac{q^{2}}{2
\pi r^{4}}\right)d r\\\nonumber&+\frac{r^{3}}{2}
\int_{r}^{r_{\Sigma}} \frac{e^\frac{{\xi -
\chi}}{2}}{r}\left(\xi^{''}+\frac{\xi^{'}}{r}+\frac{\xi^{'2}}{2}-\frac{\xi^{'}\chi^{'}}{2}\right)dr+r^{3}\int_{r}^{r_{\Sigma}}
\frac{1}{r^{2}(r+\frac{q^{2}}{r})}\left[\frac{1}{2}e^{\frac{\xi+\chi}{2}}-\frac{r^{2}}{2}\right.\\\label{62}&\left.\times
e^{\frac{\xi-\chi}{2}}\left(\frac{2
\chi^{'}}{r}+\frac{1}{r^{2}}+\frac{\xi^{'}}{r}+\frac{\xi^{'}\chi^{'}}{2}-\xi^{''}-\frac{\xi^{'2}}{2}\right)\right]dr.
\end{align}
Joining Eqs. \eqref{24} and \eqref{62} results in
\begin{align}\label{63}
\int_{r}^{r_{\Sigma}}\frac{e^{\frac{\xi+\chi}{2}}}{(r+\frac{q^{2}}{r})}\left[Y_{TF}-L_{e
\omega}+2\mathcal{T}^{2(cr)}_{2}+\frac{q^{2}}{8 \pi r^{4}}-\bigg\{4
\pi \left(P_{\bot}-P_{r}+2\mathcal{T}^{2(cr)}_{2}+\frac{q^{2}}{8 \pi
r^{4}}\right) -\in\bigg\}\right]dr=0.
\end{align}
Equation \eqref{63} implies that $Y_{TF}$ describes the importance of self-gravitational source of the intricate composition having anisotropic pressure and non-homogenous distribution of energy on Tolman mass in $f(R, L_{m}, \mathcal{T})$ gravity. Additionally, $Y_{TF}$ illustrates that these equations are of significant relevance in interpreting the varying Tolman mass expressions as compared to homogenous energy distribution and ideal fluid. Nevertheless, the Tolaman mass can also be determined through the expression given below
\begin{align}\label{64}
m_{T}=\int_{0}^{r} r^{2} e^{\frac{\xi+\chi}{2}}\left[Y_{T}-D+4
\pi\left(P_{r}+2P_{\bot}+\mathcal{T}_{1}^{1(cr)}-\mathcal{T}_{0}^{0(cr)}+2
\mathcal{T}_{2}^{2(cr)}\right)-\frac{2q^{2}}{r^{4}}\right]d
r+e^{\frac{\xi- \chi}{2}} \frac{\xi^{'} r^{2}}{2}.
\end{align}
Equation \eqref{64} directly correlates the primary variables,
Tolman mass and correction terms resulting from $f(R, L_{m},
\mathcal{T})$ gravity. It is important to mention that $Y_{T}$ has a
substantial relation with mass density. Herrera et al.
\cite{herrera2012cylindrically} and others
\cite{bhatti2017gravitational,yousaf2017role} demonstrated the
effects of $Y_{T}$ on Raychaudhuri equation, a renowned framework
for the elaboration of the expanding universe. The above-mentioned
equation specifies that Raychaudhuri equation must be described by
the consideration of $m_{T}$, irrespective of the alternative
gravity theory.

The concept of complexity in self-gravitating systems was recently
introduced by L. Herrera \cite{herrera2018new}, emphasizing the role
of energy density inhomogeneity and local anisotropy in defining the
complexity of a relativistic system. The orthogonal splitting of the
Riemann tensor allows the formulation of structure scalars, among
which $Y_{TF}$ emerges as the most significant in evaluating
complexity. This scalar encapsulates the effects of density
gradients and anisotropic pressures, making it a crucial factor in
determining the physical properties of stellar structures. In the
framework of $f(R,L_{m},\mathcal{T})$ gravity, the presence of
additional curvature-matter couplings introduces dark source terms
that can influence the anisotropy and inhomogeneity, thereby
modifying CF. By analyzing the behavior of $Y_{TF}$ and $X_{TF}$ for
various stellar models, we can assess how different modifications of
gravity affect the internal structure and stability of compact
objects.

To strengthen the validity of our model, we consider a particular
minimal model of $f(R,L_{m},\mathcal{T})$ gravity given by
\cite{haghani2021generalizing}
\begin{equation}\label{3c}
f(R,L_m,T)=R+c_0 f_1(R)+ 2c_1 f_2(L_m)+c_2 f_3(T)=R+2c_1 L_m + c_2
T,
\end{equation}
where $c_0,~c_1$ and $c_2$ are arbitrary constants having null
dimension. We further analyze the behavior of $Y_{TF}$ and $X_{TF}$
for a physically viable stellar structure. By plotting this scalar
for a particular star, namely 4U 1820-30 with the mass
$\texttt{M}=1.58 \pm 0.06 ~\texttt{M}_{\bigodot}$ and radius
$\texttt{R}=9.1 \pm 0.4 ~km$ \cite{42aa}, we establish a direct link
between our theoretical findings and astrophysical measurements. In
particular, we compare the trends of these quantities with recent
studies in GR and other extended theories such as $f(R,T)$ and
Rastall gravity frameworks. These comparisons help us to elucidate
the role of modified gravity in determining the complexity of
celestial models. Recent observational studies on compact objects
such as neutron stars, white dwarfs, and quark stars provide
constraints on mass, radius, and internal anisotropies. By utilizing
these constraints, we can benchmark our results against
well-established models. Specifically, our analysis highlights the
conditions under which $Y_{TF}$ vanishes, corresponding to systems
with minimal complexity, and the scenarios where it attains maximum
values, indicative of highly anisotropic and inhomogeneous matter
distributions.

By incorporating an analysis of the scalars $Y_{TF}$ and $X_{TF}$ in
Figure \textbf{1} using a minimal model \eqref{3c} and comparing it
with observationally motivated configurations, we provide a more
comprehensive outlook on the implications of
$f(R,L_{m},\mathcal{T})$ gravity. It must be highlighted that the
factor $Y_{TF}$ starts from zero, taking its maximum for a
particular value of the radial coordinate and then again approaches
to zero at the stellar radius. This implies that the complexity is
minimum at the center as well as the stellar surface, and achieves
its maximum otherwise. This result is in contrast with GR
\cite{paper26}, $f(R,T)$ \cite{paper15} and Rastall theories
\cite{papersl2} where CF does not vanish at the surface of a star.
Further, the factor $X_{TF}$ shows similar trend but we can't claim
this as CF because it does not fulfill the definition of CF when one
vanishes the effect of the considered modified gravity. This
approach not only strengthens the theoretical foundation of our
study but also enhances its relevance in the context of real
astrophysical systems. Thus, our study offers both a mathematical
and phenomenological perspective on CF, making it a valuable
contribution to the ongoing research on relativistic stellar models
in alternative gravitational theories.
\begin{figure}\center
\epsfig{file=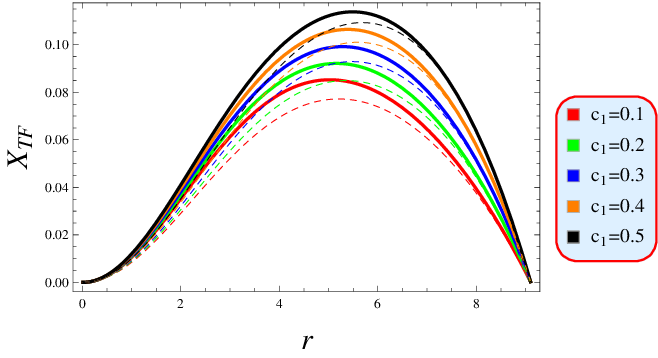,width=0.5\linewidth}\epsfig{file=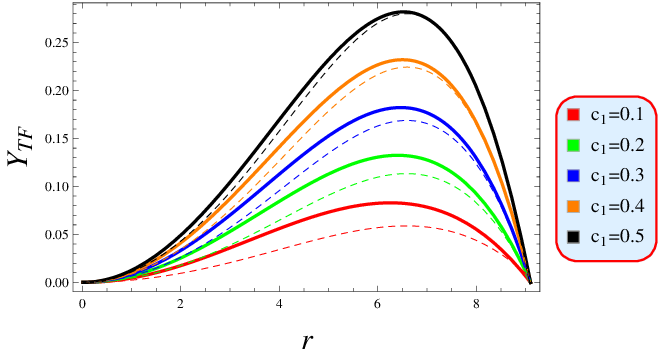,width=0.5\linewidth}
\caption{Plots of $X_{TF}$ \eqref{57} and $Y_{TF}$ \eqref{60} versus
$r$ with $c_2=0.3$ (solid) and $0.7$ (dashed).}
\end{figure}

\section{Different fluid models in the context of vanishing complexity factor}

The diminishing of CF in the framework of particular models is
assessed in this section. The assessment of complexity in different
fields of study claims that a structure scalar obtained from the
splitting of the Riemann curvature tensor is linked with the
intricacy of matter distribution. The study proposes $Y_{TF}$ as CF,
accounting for anisotropic pressure, non-homogenous energy density,
and correction terms appeared due to the $f(R, L_{m}, \mathcal{T})$
gravity. The revised equations of motion contain five variables
requiring two extra constraints to obtain a single solution. The
diminishing complexity is considered one of these restraints that
results in
\begin{align}\nonumber
\Pi&=\frac{1}{r^{3}}\int_{0}^{r}
r^{3}\left(\rho+\mathcal{T}^{0(cr)}_{0}\right)^{'}dr+\frac{q^{2}}{16
\pi r^{4}}-\frac{3L_{e \omega }}{8
\pi}-\left(P_{r}+2P_{\bot}-\mathcal{T}^{1(cr)}_{1}-2\mathcal{T}^{2(cr)}_{2}\right)\\\label{65}&-\frac{q^{2}}{16
\pi r^{3}}+\frac{q^{2}}{\pi r^{4}}+\frac{3}{2 \pi r^{3}}\int_{0}^{r}
 \frac{q^{2}}{r^{2}}dr-\frac{3}{4 \pi r^{3}}\int_{0}^{r}
 \frac{q q^{'}}{r}dr.
\end{align}
The above equation asserts that vanishing CF condition signifies
either uniform energy density and isotropic pressure or non-uniform
energy density in addition with anisotropic pressure canceling the
effects of each other. Moreover, this expression is considered as
non-local EoS in $f(R, L_{m}, \mathcal{T})$ gravity.

\subsection{Gokhroo-Mehra ansatz}

Gokhroo and Mehra \cite{gokhroo1994anisotropic} studied the inner composition of spherically symmetric matter distribution  in order to evaluate the mechanics of celestial formations. The energy density is written in following form in this scenario
\begin{equation}\label{66}
\rho=\left(1-\frac{Y r^{2}}{r_{\Sigma}^{2}}\right)
\rho_{0},
\end{equation}
where $Y$ is a constant with the value having range $(0,1)$. After
putting its value in Eq. \eqref{16}, we get
\begin{equation}\label{67}
m=\frac{4 \pi r^{3}}{3}  \left(1-\frac{3 Y r^{2}}{5
r^{2}_{\Sigma}}\right)\rho_{0}+4 \pi \int_{0}^{r} r^{2}
\mathcal{T}_{0}^{0(cr)} d r- 4 \pi \int_{0}^{r} \frac{q^{2}}{8 \pi r^{2}}dr +\int_{0}^{r} \frac{q q^{'}}{r} dr -\int_{0}^{r} \frac{q^{2}}{2r^{2}}dr .
\end{equation}
The combination of Eq. \eqref{67} and \eqref{15} concludes
\begin{align}\nonumber
e^{-\chi}&=1+\frac{q^{2}}{r^{2}}-8 \pi r^{2} \sigma\left(1-\frac{3 Y r^{2}}{5
r^{2}_{\Sigma}}\right)-\frac{8 \pi}{r} \int_{0}^{r} r^{2}
\mathcal{T}_{0}^{0(cr)} d r+\frac{8 \pi}{r} \int_{0}^{r} \frac{q^{2}}{8 \pi r^{2}}dr\\\label{68}& -\frac{2}{r}\int_{0}^{r} \frac{q q^{'}}{r}dr+\frac{2}{r}\int_{0}^{r} \frac{q^{2}}{2 r^{2}}dr .
\end{align}
Moreover, the use of second and third modified field equations yield
the the following mathematical form
\begin{align}\label{69}
8
\pi\left\{P_{r}-P_{\bot}-\mathcal{T}_{1}^{1(cr)}+\mathcal{T}_{2}^{2(cr)}+\frac{q^{2}}{4
\pi
r^{4}}\right\}=-\frac{1}{r^{2}}+e^{-\chi}\left[\frac{\xi^{\prime}}{2}+\frac{1}{r^{2}}+\xi^{\prime}
\chi^{\prime}-2 \xi^{\prime \prime}-\xi^{\prime2}+\frac{2
\chi^{\prime}}{r} -\frac{2 \xi^{\prime}}{r}\right].
\end{align}

Now, we introduce new variables having significant implications as
\begin{equation}\label{70}
e^{\xi(r)}=\frac{1}{e^{ \int_{0}^{r} \left( \frac{2}{r}-2
\varpi(r)\right) d r}},  \quad  e^{\chi  (r)}=\frac{1}{g}.
\end{equation}
Equation \eqref{69}  takes the subsequent form in the context of
consideration of above new variables
\begin{equation}\label{71}
g^{\prime}-g\left[\frac{6}{2 r}-\frac{5}{2r^{2} \varpi}-\frac{2
\varpi^{\prime}}{\varpi}-2 \varpi\right]= \frac{4
\pi}{\varpi}\left[\mathcal{T}_{1}^{1(cr)}-\mathcal{T}_{2}^{2(cr)}-\Pi-\frac{q^{2}}{4
\pi r^{4}}-\frac{1}{r^{2}}\right].
\end{equation}
This mathematical form is similar to Ricatti's equation and is based
on assessing the $\chi$ function associated with the line element,
retained in $g(r)$ and described in Eq. \eqref{68} in
addition to the expression for $\Pi$ given in \eqref{65}. After
the contemplation of the integral form of the preceding equation,
the metric can be stated as dependent to $\varpi$ and $\Pi$ within
the context of $f(R, L_{m}, \mathcal{T}$) gravity. Consequently, we
can write
\begin{eqnarray}\nonumber
&&d s^{2}=e^{2 \int_{0}^{r} \left(\varpi-\frac{1}{r}\right) d r} d
t^{2}-r^{2}\left(d^{2} \theta+\sin ^{2} \theta d^{2}
\phi\right)\\\label{72}&&-\frac{\varpi^{2} e^{2 \int_{0}^{r}
\left(\varpi+\frac{5}{2r^{2} \varpi}\right) d r}}{4 \pi r^{6}
\int_{0}^{r}
\frac{\varpi}{r^{6}}\left[\frac{1}{r^{2}}+\Pi+\mathcal{T}_{2}^{2(cr)}-\mathcal{T}_{1}^{1(cr)}\right]
e^{2 \int_{0}^{r} \left(\varpi+\frac{5}{2r^{2}\varpi}\right) d r} d
r+U}dr^2,
\end{eqnarray}
where $U$ is the integration constant. Furthermore, the primary
parameters are modified in subsequent form as a result of insertion of new
variables
\begin{align}\label{73}
4 \pi\rho&=-4\pi\mathcal{T}_{0}^{0(cr)}+\frac{m^{\prime}}{r^{2}}+\frac{q^{2}}{2 r^{4}}-\frac{q q^{'}}{r}+\frac{q^{2}}{2 r^{2}},
\\\label{74}
4 \pi P_{r}&=4 \pi \mathcal{T}_{1}^{1(cr)}-\frac{q^{2}}{2
r^{4}}-\frac{1}{2
r^{2}}+\left(\frac{2r\varpi-1}{2r^{2}}\right)\left(1-\frac{2m}{r}+\frac{q^{2}}{r^{2}}\right),
\\\nonumber
8 \pi P_{\bot}&=8\pi \left(\mathcal{T}_{2}^{2(cr)}-\frac{q^{2}}{8
\pi r^{4}}\right)-\frac{\left(1-\frac{2
m}{r}+\frac{q^{2}}{r^{2}}\right)}{4}\\\label{75}&\times\left[\frac{2
g^{'}}{rg}-\frac{2\varpi
g^{'}}{g}-4\varpi^{'}-\frac{4}{r^{2}}-4\left(\varpi-\frac{1}{r}\right)^{2}-\frac{2g^{'}}{r
g}-\frac{4\varpi}{r}+\frac{4}{r^{2}}\right].
\end{align}
The positive energy density in addition to the restriction $ \rho >
P_{r}, P_{\bot}$ suggests that the development of important
techniques to clarifying the gravitational structure are consistent.
Moreover, these equations can be helpful in explaining some
intriguing and fascinating aspects of the system. The establishment of singularities due to contemplated variables
corresponding to the hypersurface are avoided after the
consideration of Darmois restraints within the context of
appropriate external solutions associated with the internal configuration
of the structure.

\subsection{Polytropic equation of state}

The polytropic EoS implies significant implications for understanding the
configuration of self-gravitating structure. In this manuscript, the
polytropic composition is considered after assuming the restraint of
vanishing CF. Following that, we consider two certain polytropic
schemes
\cite{herrera2013newtonian,herrera2013general,herrera2016cracking}.
We proceed in the following way
\begin{equation}\label{76}
P_{r}=\mathcal{K}\rho^{\tau}=\mathcal{K}\rho^{1+\frac{1}{n}},
\quad \tau=1+\frac{1}{n}.
\end{equation}
where $\mathcal{K}$ is defined as polytropic constant whereas,
$\tau$ and $n$ specify the polytropic exponent and index,
respectively. It is easy to solve an equation in dimensionless form
so, we introduce more parameters given below for the dimensionless
formulations of the TOV equation and the mass function. They are
\begin{equation}\label{77}
\psi=\frac{P_{rc}}{\rho_{c}}, \quad r=\frac{\mu}{D}, \quad
D^{2}=\frac{4 \pi \rho_{c}}{\psi(n+1)}, \quad
\vartheta(\mu)=\frac{D^{3}m(r)}{4 \pi \rho_{c}}, \quad
\tau^{n}=\frac{\rho}{\rho_{c}},
\end{equation}
in which subscript $c$ indicates the execution of the quantity at
the center. We contemplate $\tau(\mu_{\Sigma})=0$ at the horizon
$r=r_{\Sigma}(\mu=\mu_{\Sigma})$. The TOV equation in terms of these
dimensionless parameters is expressed in the subsequent form
\begin{align}\nonumber
&2\mu^{2} \frac{d \tau}{d \mu}\left[\frac{1-2 \psi(n+1)
\vartheta / \mu+\frac{q^{2}D}{2 \mu}}{1+\psi
\tau+\frac{e^{\xi}T^{00(D)}}{\tau^{n}\rho_{c}}+\frac{q^{2}\rho_{c}}{4 \pi \psi^{2}(n+1)^{2}\tau^{n}}}\right]+16
\pi \vartheta+8 \pi \psi \mu^{3} \tau^{1+n} -\frac{8 \pi
\mu^{3}}{\rho_{c}} \mathcal{T}_{1}^{1(D)}\\\nonumber&-\frac{q^{2}D}{4 \psi (n+1)}+\left[\frac{D\left(1-\frac{2\psi(n+1)\vartheta}{\mu}+q^{2}\right)\left(1-\frac{2\psi(n+1)\vartheta}{\mu}+\frac{q^{2} D}{\mu}\right)}{\left(P_{r
c}\tau^{n}(n+1)\right)\left(1+\psi \tau+\frac{e^{\xi}\mathcal{T}^{00(D)}}{\tau^{n}\rho_{c}}+\frac{\rho_{c}q^{2}}{4 \pi\mu^{4}\tau^{n}}\psi^{2}(n+1)^{2}\right)}\right]\\\label{78}& \times \left[\mathcal{T}^{11'(cr)}+\chi^{'}\mathcal{T}^{11(cr)}+\frac{2D}{\mu}\mathcal{T}^{11(cr)}-\frac{\mu}{D e^{\chi}}\mathcal{T}^{22(cr)}+\Pi-Z^{\star}+\left(\frac{3 q^{2}\rho^{2}_{c}}{8 \pi \mu^{4}e^{\chi}\psi^{2}(n+1)^{2}}\right)^{'}\right]=0.
\end{align}
Also, Eq. \eqref{16} after the consideration of above
mentioned dimensionless variables becomes
\begin{equation}\label{79}
\frac{d \vartheta}{d \mu}=4 \pi
\mu^{2}\left(\tau^{n}+\frac{\mathcal{T}^{0(cr)}_{0}}{\rho_{c}}\right)-\frac{q^{2}\rho_{c}}{2 \mu^{2}\psi^{2}(n+1)^{2}}+\frac{q q^{'}\rho_{c}}{\mu \psi^{2}(n+1)^{2}}.
\end{equation}

The appearance of $\Pi$, $\vartheta$, and $\tau$ functions in above
two equations point out the requirement of one additional constraint
to obtain a definite solution related to the system. Consequently,
we write the disappearing CF constraint in dimensionless variables
in the following form
\begin{align}\nonumber
\frac{2 \mu}{n \rho_{c}}\frac{d \Pi}{d \mu}+\frac{6 \Pi}{n \rho_{c}}&=\frac{\mu}{4 \pi n \rho_{c}}\frac{d}{d \mu}\left(\frac{q^{2}\rho_{c}}{\mu^{4}\psi^{2}(n+1)^{2}}\right)-\frac{3 \mu}{8 \pi n\rho_{c}}\frac{d}{d \mu}L_{e \psi}+\frac{r}{n \rho_{c}}\left(\rho+\mathcal{T}^{0(cr)}_{0}\right)^{'}\\\nonumber&+\frac{\mu}{n \rho_{c}}\frac{d}{d\mu}\left(\mathcal{T}^{1(cr)}_{1}+2\mathcal{T}^{2(cr)}_{2}\right)-\frac{3 \mu}{16 \pi n\rho_{c}}\frac{d}{d \mu}\left(\frac{q^{2}\rho_{c}}{r \mu^{2}\psi (n+1)}\right)\\\nonumber&+\frac{9 q^{2}\rho_{c}}{4 \pi \mu^{4}n \psi^{2}(n+1)^{2}}
-\frac{3 q q^{'}}{4 \pi r \mu^{2}\psi n(n+1)}+\frac{9 \rho_{c}}{4 \pi \mu^{3}\psi^{2}n^{2}(n+1)^{2}}\\\nonumber& \times
\int \frac{q q^{'}}{r}dr-\frac{9}{8 \pi n \rho_{c}}+\frac{3}{n \rho_{c}}\int \left(\rho+\mathcal{T}^{0(cr)}_{0}\right)^{'}dr
-\frac{3}{n \rho_{c}}\left(\mathcal{T}^{1(cr)}_{1}+2\right.\\\label{80}& \times \left.\mathcal{T}^{2(cr)}_{2}\right)-\frac{9 q^{2}}{16 \pi r \mu^{2}\psi n(n+1)}-\frac{9}{4 \pi r \mu^{2}\psi n(n+1)}\int \frac{q q^{'}}{r}dr.
\end{align}
The interpretive solution of these equations
can be linked with various values of $\vartheta$ and $n$, or their
numeric solutions can be obtained by considering pertinent
restraints. Each solution can estimate mass, density, and pressure
for cosmological formations relying on independent variables.

Moreover, the second polytropic EoS might be regarded as
\begin{equation}\label{76b}
P_{r}=\mathcal{K}\rho_{d}^{\tau}=\mathcal{K}\rho_{d}^{1+\frac{1}{n}},
\quad \tau=1+\frac{1}{n}.
\end{equation}
in which $\rho_{d}$ is the baryonic mass density. The consideration
of above equation in addition to \eqref{78} and \eqref{80} results into
\begin{align}\nonumber
&2\mu^{2} \frac{d \tau_{d}}{d \mu}\left[\frac{1-2 \psi(n+1)
\vartheta / \mu+\frac{q^{2}D}{2 \mu}}{1+\psi
\tau_{d}+\frac{e^{\xi}T^{00(D)}}{\tau^{n}_{d}\rho_{cd}}+\frac{q^{2}\rho_{cd}}{4
\pi \psi^{2}(n+1)^{2}\tau^{n}_{d}}}\right]+16 \pi \vartheta+8 \pi
\psi \mu^{3} \tau^{1+n}_{d} -\frac{8 \pi \mu^{3}}{\rho_{cd}}
\mathcal{T}_{1}^{1(D)}\\\nonumber&-\frac{q^{2}D}{4 \psi
(n+1)}+\left[\frac{D\left(1-\frac{2\psi(n+1)\vartheta}{\mu}+q^{2}\right)\left(1-\frac{2\psi(n+1)\vartheta}{\mu}+\frac{q^{2}
D}{\mu}\right)}{\left(P_{r c}\tau^{n}_{d}(n+1)\right)\left(1+\psi
\tau_{d}+\frac{e^{\xi}\mathcal{T}^{00(D)}}{\tau^{n}_{d}\rho_{cd}}+\frac{\rho_{cd}q^{2}}{4
\pi\mu^{4}\tau^{n}_{d}}\psi^{2}(n+1)^{2}\right)}\right]\\\label{78b}&
\times
\left[\mathcal{T}^{11'(cr)}+\chi^{'}\mathcal{T}^{11(cr)}+\frac{2D}{\mu}\mathcal{T}^{11(cr)}-\frac{\mu}{D
e^{\chi}}\mathcal{T}^{22(cr)}+\Pi-Z^{\star}+\left(\frac{3
q^{2}\rho^{2}_{cd}}{8 \pi
\mu^{4}e^{\chi}\psi^{2}(n+1)^{2}}\right)^{'}\right]=0,
\end{align}
and
\begin{align}\nonumber
\frac{2 \mu}{n \rho_{cd}}\frac{d \Pi}{d \mu}+\frac{6 \Pi}{n \rho_{cd}}&=\frac{\mu}{4 \pi n \rho_{cd}}\frac{d}{d \mu}\left(\frac{q^{2}\rho_{cd}}{\mu^{4}\psi^{2}(n+1)^{2}}\right)-\frac{3 \mu}{8 \pi n\rho_{cd}}\frac{d}{d \mu}L_{e \psi}+\frac{r}{n \rho_{cd}}\left(\rho_{d}+\mathcal{T}^{0(cr)}_{0}\right)^{'}\\\nonumber&+\frac{\mu}{n \rho_{cd}}\frac{d}{d\mu}\left(\mathcal{T}^{1(cr)}_{1}+2\mathcal{T}^{2(cr)}_{2}\right)-\frac{3 \mu}{16 \pi n\rho_{cd}}\frac{d}{d \mu}\left(\frac{q^{2}\rho_{cd}}{r \mu^{2}\psi (n+1)}\right)\\\nonumber&+\frac{9 q^{2}\rho_{cd}}{4 \pi \mu^{4}n \psi^{2}(n+1)^{2}}
-\frac{3 q q^{'}}{4 \pi r \mu^{2}\psi n(n+1)}+\frac{9 \rho_{cd}}{4 \pi \mu^{3}\psi^{2}n^{2}(n+1)^{2}}\\\nonumber& \times
\int \frac{q q^{'}}{r}dr-\frac{9}{8 \pi n \rho_{cd}}+\frac{3}{n \rho_{cd}}\int \left(\rho_{d}+\mathcal{T}^{0(cr)}_{0}\right)^{'}dr
-\frac{3}{n \rho_{cd}}\left(\mathcal{T}^{1(cr)}_{1}+2\right.\\\label{80b}& \times \left.\mathcal{T}^{2(cr)}_{2}\right)-\frac{9 q^{2}}{16 \pi r \mu^{2}\psi n(n+1)}-\frac{9}{4 \pi r \mu^{2}\psi n(n+1)}\int \frac{q q^{'}}{r}dr,
\end{align}
where $\tau_{d}^{n}=\frac{\rho_{d}}{\rho_{cd}}$. The development of
compact formations can be analyzed by solving Eqs. \eqref{79},
\eqref{78b}, and \eqref{80b} under the constraints mentioned in
\eqref{76} along with the diminishing complexity. The polytropic EoS
in addition to radial component of pressure and energy density can
play a crucial role in characterizing distinct phases of expanding
universe. The prevalence of matter, radiation, and stiff fluid are
specified through $\mathcal{K}=0, \frac{1}{3}$, and $1$.
Furthermore, the phantom era is illustrated after the consideration
of $\mathcal{K} < -1$, whereas $\mathcal{K}\in (\frac{-1}{3}, -1)$
evaluates the quintessence period. Different values of polytropic
index $n$ are important in illustrating distinct matter
configurations related to the compact objects, i.e., $n=0.5$ and
$n=1$ are considered the appropriate selection for the
description of the composition of neutron stars.

\section{Implications of our findings in the
context of stellar structures}

The study of complexity in relativistic systems plays a crucial role
in understanding the internal structure and evolution of compact
stellar objects such as neutron stars and quark stars. In our
analysis, we employ the orthogonal splitting of the Riemann
curvature tensor to explore the implications of complexity in the
context of $f(R,L_{m},\mathcal{T})$ gravity. Below, we outline the
practical significance of our findings and their broader relevance
to astrophysics and gravitational physics.
\begin{itemize}
\item \textbf{Understanding compact stellar structures:} The decomposition of
the curvature tensor into its fundamental components provides an
avenue to measure the deviation from homogeneity within
self-gravitating systems. Our findings indicate that the presence of
anisotropies and inhomogeneous energy distributions contribute
significantly to the complexity of the system. This understanding is
critical in modeling neutron stars and strange stars, where
anisotropic pressures are prevalent due to extreme densities and
strong gravitational fields.
\item \textbf{Improved EoS constraints:} The evaluation of
complexity through the scalar $Y_{TF}$ allows for a more refined
approach to constructing EoSs for compact objects. Since different
astrophysical observations suggest varying degrees of anisotropy in
neutron stars, our framework provides theoretical constraints that
could guide the development of more realistic EoS models. These
models are essential for predicting mass-radius relationships and
stability conditions of compact stars.
\item \textbf{Influence of modified gravity on stellar configurations:} The
framework of $f(R,L_{m},\mathcal{T})$ gravity introduces additional
dark source terms, which have direct implications for the mass and
pressure profiles of compact objects. By linking the Tolman and
Misner-Sharp mass formulations with the conformal tensor, we have
provided a mechanism to evaluate how modifications to GR impact the
structure of relativistic objects. This insight is particularly
relevant in exploring alternative gravity theories as viable
explanations for observed astrophysical phenomena, such as dark
matter effects and deviations from the standard GR predictions.
\item \textbf{Astrophysical observations and gravitational wave astronomy:} The
study of complexity in stellar interiors is increasingly relevant in
the era of multi-messenger astrophysics. Gravitational wave
observations from binary neutron star mergers (e.g., GW170817)
provide constraints on the internal composition of neutron stars.
Our work suggests that the CF, influenced by anisotropies and
modified gravity terms, could have observable effects on tidal
deformability and gravitational wave signatures. Future
gravitational wave detections, combined with electromagnetic
observations, could provide an empirical test for our theoretical
predictions.
\item \textbf{Collapse dynamics and black hole formation:} The role of
complexity in the stability and evolution of compact objects has
implications for their ultimate fate. Systems with high complexity,
as characterized in our study, may be more susceptible to dynamical
instabilities leading to gravitational collapse. This understanding
is critical in predicting the conditions under which a neutron star
transitions into a black hole, as well as the formation of exotic
compact objects in alternative gravity scenarios.
\item \textbf{Thermal and electromagnetic emissions:} The presence of charge in
our analysis adds another dimension to the astrophysical relevance
of our findings. Highly magnetized neutron stars, such as magnetars,
exhibit strong electromagnetic emissions due to their extreme
magnetic fields. The relationship between complexity and charge in
our framework provides insights into the role of charge
distributions in influencing thermal and radiation processes in
compact objects.
\end{itemize}

Our findings provide a theoretical foundation for analyzing the
complexity of compact stellar structures in the context of modified
gravity. The practical implications extend to astrophysical
modeling, gravitational wave astronomy, and the study of stellar
evolution. By incorporating our results into observational
frameworks, future research can further validate the predictions
made in this work, thereby advancing our understanding of
relativistic stellar structures in both GR and extended gravity
theories.

\section{Discussion and Conclusions}

Our study interprets a particular methodology suggested by Herrera
for determining the complexity of a system in $f(R, L_{m},
\mathcal{T})$ theory. This modified form of GR more precisely
describes the fundamental physics associated with the rapid cosmic
expansion. Initially, we discussed the primary mathematical
formation of $f(R, L_{m}, \mathcal{T}$) gravity in addition to the
important parameters for the explanation of the anisotropic
dispersion of matter. The revised field equations related to
spherically symmetric geometry are then calculated. Furthermore, the
contemplation of non-conservation law results in the derivation of
the TOV equation that plays a crucial role in our further
analysis of complexity. The formalism of the considered modified
gravity is also used to formulate specific junction conditions for
smooth conjunction between the exterior region and the interior
composition of the charged fluid. After the consideration of spherical
charged anisotropic composition of matter, we have considered the
formulas of Misner-Sharp and Tolman masses that ultimately
contributed in determining the relation between conformal tensor and
other basic parameters. The significance of conformal tensor,
Misner-Sharp mass, electric charge and the gravitational mass on the
matter dispersion linked with the anisotropic distribution of
non-homogenous energy density and the dark source terms relating to
$f(R, L_{m}, \mathcal{T})$ gravity is also assessed. The orthogonal
decomposition of the Riemann curvature tensor resulted in the
derivation of $Y_{TF}$ as the CF because it incorporates the
anisotropic pressure, electric charge and non-homogenous energy
density. The behavior of fluctuating energy density proposed by
Gokhroo and Mehra \cite{gokhroo1994anisotropic} is evaluated from the
perspective of vanishing complexity. Finally, the polytropic EoS as
well as the null complexity in the presence of electric charge is
studied to derive the feasible solutions for the spherically
symmetric structure. We wrap up the discussion with the following
comments.
\begin{itemize}
\item Several factors, particularly the anisotropy in the matter
configuration, the presence of electric charge, and the
non-homogenous energy density play a crucial role in
evaluating the complexity of any structure. The structural variables
proposed by Herrera \cite{herrera2009structure} are determined
after the orthogonal splitting of curvature tensor. We
analyzed the five structural variables along with the investigation
of their related features and determined solutions within the
framework of these parameters.

\item The quantities $Y_{T}$  and $Y_{TF}$ are categorized as the key
competitors for explaining the development of shear and expansion
tensors. The $Y_{TF}$ factor accounts for the influence of
homogenous energy density, electric field, modified dark source
terms, and anisotropic composition of matter on energy dispersion.
The consistency of vanishing complexity constraint is maintained by
contemplating certain frameworks to evaluate the system.

\item The energy density proposed by Gokhroo and Mehra
is used as a seed solution in the framework of $f(R, L_{m},
\mathcal{T}$) theory. Some dimensionless variables are also
introduced which help in solving the equations more easily.

\item The TOV equation, diminshing CF, and
mass relations are derived after the assumption of polytropic EoS.
It is critical to note that the consideration of isotropic pressure
and uniform energy density results in the disappearance of CF in the
context of GR \cite{herrera2018new}. Despite of such constraints,
findings acquired in this manuscript are compatible with the vanishing complexity,
demonstrating the significance of correction terms relating to the
$f(R, L_{m}, \mathcal{T}$) gravity.
\end{itemize}

The findings of our study might be employed for the precise
understanding of significant astronomical events. Following this
effort, substantial models can be developed for the thorough
investigation of relation between $f(R, L_{m}, \mathcal{T}$)
gravity and the accelerating cosmic expansion.

\section*{Appendix}

\renewcommand{\theequation}{A\arabic{equation}}
\setcounter{equation}{0} \noindent The correction terms appearing in Eqs. \eqref{9}-\eqref{11} are
\begin{align}\nonumber
\mathcal{T}^{00(cr)}&=\frac{f_{\mathcal{T}}e^{\xi}\rho}{8 \pi
f_{R}}+\frac{f_{L_{m}}e^{\xi}\rho}{16 \pi
f_{R}}+\frac{\xi^{'}e^{\xi-\chi}}{16 \pi
f_{R}}f_{R}^{'}+\frac{e^{\xi-\chi}}{8 \pi
f_{R}}f_{R}^{''}-\frac{\chi^{'}e^{\xi-\chi}}{16 \pi
f_{R}}f_{R}^{'}+\frac{e^{\xi-\chi}}{4 \pi
f_{R}}f_{R}^{'}-\frac{\xi^{'}e^{\xi-\chi}}{16 \pi
f_{R}}\\\nonumber& \times f_{R}^{'}+\frac{f e^{\xi}}{16 \pi
f_{R}}-\frac{R f_{R}}{16 \pi
f_{R}}e^{\xi}-\frac{f_{\mathcal{T}}L_{m}e^{\xi}}{4
\pi f_{R}}-\frac{f_{L_{m}}L_{m}e^{\xi}}{8
\pi f_{R}},
\\\nonumber
\mathcal{T}^{11(cr)}&=\frac{e^{\chi}P_{r}f_{\mathcal{T}}}{8 \pi
f_{R}}+\frac{e^{\chi}P_{r}f_{L_{m}}}{16 \pi
f_{R}}-\frac{\xi ^{'}}{16 \pi
f_{R}}f_{R}^{'}-\frac{f_{R}^{''}}{8 \pi f_{R}}+\frac{\chi^{'}}{16
\pi f_{R}}f_{R}^{'}-\frac{f_{R}^{'}}{4 \pi r
f_{R}}+\frac{f_{R}^{''}}{8 \pi
f_{R}}\\\nonumber&-\frac{\chi^{'}f_{R}^{'}}{16 \pi f_{R}}-\frac{f
e^{\chi}}{16 \pi f_{R}}+\frac{R f_{R}e^{\chi}}{16 \pi
f_{R}}+\frac{e^{\chi}f_{\mathcal{T}}L_{m}}{4 \pi
f_{R}}+\frac{e^{\chi}f_{L_{m}}L_{m}}{8 \pi
f_{R}},
\\\nonumber
\mathcal{T}^{22(cr)}&=\frac{f_{\mathcal{T}}r^{2}P_{\bot}}{8 \pi
f_{R}}+\frac{f_{L_{m}}r^{2}P_{\bot}}{16 \pi f_{R}}
-\frac{\xi^{'}r^{2}e^{-\chi}}{16 \pi
f_{R}}f_{R}^{'}-\frac{r^{2}e^{-\chi}}{8 \pi
f_{R}}f_{R}^{''}+\frac{r^{2}e^{-\chi}\chi^{'}}{16 \pi
f_{R}}f_{R}^{'}-\frac{e^{-\chi}r}{4 \pi f_{R}}f_{R}^{'}
\\\nonumber& +\frac{r e^{-\chi}f_{R}^{'}}{8 \pi f_{R}}-\frac{r^{2}f}{16 \pi f_{R}}+\frac{r^{2}R f_{R}}{16 \pi f_{R}}+\frac{r^{2}L_{m}f_{\mathcal{T}}}{4 \pi f_{R}}+\frac{r^{2}L_{m}f_{L_{m}}}{8 \pi f_{R}}.
\end{align}

\noindent The expression for $Z^{\star}$ is
\begin{align}\nonumber
Z^{\star}&=\frac{1}{8 \pi
+f_{\mathcal{T}}+\frac{1}{2}f_{L}}\left[\frac{1}{6}\left(2
P_{r}f_{\mathcal{T}}^{'}+P_{r}f_{L}^{'}+4
P_{\bot}f_{\mathcal{T}}^{'}+2
P_{\bot}f_{L}^{'}+2f_{\mathcal{T}}P_{r}^{'}+4
f_{\mathcal{T}}P_{\bot}^{'}+f_{L}P_{r}^{'}\right.\right.\\\nonumber&\left.\left.+2f_{L}P_{\bot}^{'}\right)
+e^{2 \chi}P_{r}f_{\mathcal{T}}^{'}+\frac{e^{2
\chi}}{2}P_{r}f_{L}^{'}-\frac{f_{\mathcal{T}}\rho^{'}}{2}+\frac{f_{\mathcal{T}}P_{r}^{'}}{2}+f_{\mathcal{T}}P_{\bot}^{'}-\frac{f_{L}P_{r}^{'}}{6}
-\frac{P_{\bot}^{'}}{3}\right].
\end{align}

\noindent The term $\mathcal{F}_{0}$ mentioned in Eq. \eqref{D4} is
\begin{align}\nonumber
\mathcal{F}_{0}&=-\frac{e^{2 \chi}P_{r}f_{\mathcal{T}}}{8 \pi f_{R}}
-\frac{e^{2 \chi}P_{r}f_{L_{m}}}{16 \pi
f_{R}}+\frac{\xi^{'}e^{\chi}}{16 \pi f_{R}}f_{R}^{'}
+\frac{f_{R}^{''}}{8 \pi f_{R}}-\frac{\chi^{'}e^{\chi}f_{R}^{'}}{16
\pi f_{R}} +\frac{e^{\chi}f_{R}^{'}}{4 \pi r
f_{R}}-\frac{e^{\chi}}{8 \pi f_{R}}f_{R}^{''}\\\nonumber&
+\frac{\chi^{'}e^{\chi}}{16 \pi f_{R}}f_{R}^{'}+\frac{f e^{2
\chi}}{16 \pi f_{R}} -\frac{R f_{R}e^{2 \chi}}{16 \pi
f_{R}}-\frac{e^{2 \chi}f_{\mathcal{T}}L_{m}}{4 \pi f_{R}}
-\frac{e^{2 \chi}f_{L_{m}}L_{m}}{8 \pi f_{R}}-\frac{3 Q^{2}}{8 \pi
r^{4}} -\frac{1}{8 \pi r^{2}}+\frac{1}{8 \pi}\\\nonumber&
\times\left(\frac{\xi^{'}}{r}
+\frac{1}{r^{2}}\right)e^{-\chi}\bigg|_{r=r_\Sigma}.
\end{align}
The expressions for correction terms associated with Eqs.
\eqref{53}-\eqref{58} are
\begin{align}\nonumber
\mathcal{N}_{e \omega}^{(A)}&=\left[\frac{f_{\mathcal{T}}}{32 \pi
f_{R}}\mathcal{T}^{\zeta(m)}_{\eta}+\frac{f_{L_{m}}}{64
\pi f_{R}}\mathcal{T}^{\zeta(m)}_{\eta}-\frac{\Box f_{R}}{64 \pi
f_{R}}+ \frac{f}{128 \pi f_{R}}-\frac{R f_{R}}{128 \pi
f_{R}}-\frac{f_{\mathcal{T}}L_{m}}{64 \pi
f_{R}}\right.\\\nonumber&\left.-\frac{f_{L_{m}}L_{m}}{64
\pi f_{R}} \right]\varepsilon^{\zeta \tau}_{e}\varepsilon_{\zeta
\tau \omega}+\frac{\nabla ^{\zeta}\nabla _{\eta}f_{R}}{64 \pi
f_{R}}\varepsilon ^{\eta \tau}_{e}\varepsilon_{\zeta \tau
\omega}+\left[\frac{R}{128 \pi f_{R}}-\frac{f_{\mathcal{T}}}{64 \pi
f_{R}}\mathcal{T}^{\zeta
(m)}_{\tau}-\frac{f_{L_{m}}}{128 \pi
f_{R}}\mathcal{T}^{\zeta
}_{\tau}\right.\\\nonumber&\left.+\frac{\Box f_{R}}{64 \pi
f_{R}}-\frac{\nabla ^{\zeta}\nabla _{\tau}f_{R}}{64 \pi
f_{R}}-\frac{f}{128 \pi
f_{R}}+\frac{f_{\mathcal{T}}L_{m}}{32 \pi
f_{R}}+\frac{f_{L_{m}}}{64 \pi }
\frac{L_{m}}{f_{R}}-\frac{f_{\mathcal{T}}g^{\mu \nu}}{32
\pi f_{R}}\frac{\partial ^{2}L_{m}}{\partial
g^{\zeta}_{\tau}\partial g_{\mu \nu}}\right]\varepsilon^{\eta
\tau}_{e}\varepsilon_{\zeta \eta
\omega}\\\nonumber&+\left[-\frac{f_{\mathcal{T}}}{64 \pi
f_{R}}\mathcal{T}^{\upsilon(m)}_{\eta}-\frac{f_{L_{m}}}{128
\pi
f_{R}}\mathcal{T}^{\upsilon(m)}_{\eta}-\frac{\nabla^{\upsilon}\nabla_{\eta}f_{R}}{64
\pi f_{R}} -\frac{f_{\mathcal{T}}g^{\mu \nu}}{32 \pi
f_{R}}\frac{\partial ^{2}L_{m}}{\partial
g^{\upsilon}_{\eta}\partial g_{\mu \nu}}\right]\varepsilon^{\eta
\tau}_{e}\varepsilon_{\tau \upsilon
\omega}+\left[\frac{f_{\mathcal{T}}}{32 \pi
}\right.\\\nonumber&\left.\times
\frac{\mathcal{T}^{\upsilon(m)}_{\tau}}{f_{R}}+\frac{f_{L_{m}}}{64
\pi f_{R}} \mathcal{T}^{\upsilon(m)}_{\tau}-\frac{\Box f_{R}}{64 \pi
f_{R}}+\frac{\nabla^{\upsilon}\nabla_{\tau}f_{R}}{64 \pi
f_{R}}+\frac{f}{128 \pi f_{R}}-\frac{R f_{R}}{128 \pi
f_{R}}-\frac{f_{\mathcal{T}}L_{m}}{32 \pi
f_{R}}-\frac{f_{L_{m}}}{64
\pi}\right.\\\nonumber&\left.\times \frac{L_{m}}{f_{R}}
+\frac{f_{\mathcal{T}}g^{\mu \nu}}{32 \pi
f_{R}}\frac{\partial^{2}L_{m}}{\partial
g^{\upsilon}_{\tau}\partial g_{\mu \nu}}\right]\varepsilon^{\eta
\tau}_{e}\varepsilon_{\eta \upsilon \omega},
\\\nonumber
\mathcal{N}_{e \omega}^{(B)}&=\frac{f_{\mathcal{T}}}{8 \pi
f_{R}}\mathcal{T}_{e \omega}^{(m)}+\frac{f_{L_{m}}}{16 \pi
f_{R}}\mathcal{T}_{e \omega}^{(m)}-\frac{g_{e \omega}}{8 \pi
f_{R}}\Box f_{R}+\frac{1}{16 \pi
f_{R}}\nabla_{e}\nabla_{\omega}f_{R}+\frac{f}{16 \pi f_{R}}g_{e
\omega}\\\nonumber&-\frac{R g_{e \omega}}{16 \pi
}-\frac{f_{\mathcal{T}}L_{m}}{4 \pi f_{R}}g_{e
\omega}-\frac{f_{L_{m}}L_{m}}{8 \pi f_{R}}g_{e
\omega}+\frac{f_{\mathcal{T}}g^{\mu \nu}g_{e \zeta}}{8 \pi f_{R}}
\frac{\partial^{2}L_{m}}{\partial g^{\zeta}_{\omega}\partial g_{\mu
\nu}}-\frac{u_{\omega}u^{\tau}}{8 \pi
f_{R}}f_{\mathcal{T}}\mathcal{T}^{(m)}_{e
\tau}-\frac{u_{\omega}u^{\tau}}{16 \pi
f_{R}}f_{L_{m}}\\\nonumber&\times \mathcal{T}^{(m)}_{e \tau}
+\frac{u_{\omega}u^{\tau}}{16 \pi f_{R}}g_{e \tau}\Box
f_{R}-\frac{u_{\omega}u^{\tau}\nabla_{e}\nabla_{\tau}}{16 \pi
f_{R}}f_{R}-\frac{u_{\omega}u^{\tau}}{64 \pi f_{R}}g_{e
\tau}+\frac{u_{\omega}u^{\tau}R f_{R}}{32 \pi f_{R}}g_{e
\tau}+\frac{u_{\omega}u^{\tau}}{8 \pi
f_{R}}f_{\mathcal{T}}L_{m}\\\nonumber&\times g_{e
\tau}+\frac{u_{\omega}u^{\tau}}{16 \pi f_{R}}f_{L_{m}}L_{m}g_{e
\tau} -\frac{u_{\omega}u^{\tau}}{8 \pi f_{R}}f_{\mathcal{T}}g^{\mu
\nu}g_{e \zeta} \frac{\partial ^{2}L_{m}}{\partial
g^{\zeta}_{\tau}\partial g_{\mu \nu}}-\frac{u_{\upsilon}u_{e}}{8 \pi
f_{R}}f_{\mathcal{T}}\mathcal{T}^{\upsilon(m)}_{\omega}-\frac{u_{\upsilon}u_{e}}{16
\pi f_{R}}f_{L_{m}}\mathcal{T}^{\upsilon(m)}_{\omega}
\\\nonumber&+\frac{u_{\omega}u_{e}}{16 \pi f_{R}}\Box f_{R}-\frac{u_{\upsilon}u_{e}}{16 \pi f_{R}}\nabla^{\upsilon}\nabla_{\omega}f_{R}-\frac{f u_{\omega}u_{e}}{32 \pi f_{R}}+\frac{ u_{\omega}u_{e}}{32 \pi f_{R} }R f_{R}+\frac{ u_{\omega}u_{e}}{8 \pi f_{R}}f_{\mathcal{T}}L_{m}+\frac{u_{e}u_{\omega}}{16 \pi f_{R}}f_{L_{m}}L_{m}\\\nonumber&-\frac{u_{\upsilon}u_{e}}{8 \pi f_{R}}f_{\mathcal{T}}g^{\mu \nu}\frac{\partial^{2}L_{m} }{\partial g^{\upsilon}_{\omega}\partial g_{\mu \nu}}+\frac{u_{\upsilon}u^{\tau}f_{\mathcal{T}}}{16 \pi f_{R}}g_{e\omega}\mathcal{T}^{\upsilon(m)}_{\tau}+\frac{u_{\upsilon}u^{\tau}g_{e\omega}}{32 \pi f_{R}}f_{L_{m}}\mathcal{T}^{\upsilon(m)}_{\tau}+\frac{g_{e \omega}u_{\upsilon}u^{\tau}}{16 \pi f_{R}}\nabla^{\upsilon}\nabla_{\tau}f_{R}\\\nonumber&+\frac{g_{e \omega}u_{\upsilon}u^{\tau}f_{\mathcal{T}}}{16 \pi f_{R}} \frac{\partial^{2}L_{m}}{\partial g^{\upsilon}_{\tau}\partial g_{\mu \nu}}g^{\mu
\nu},
\\\nonumber
\mathcal{N}_{e
\omega}^{(C)}&=-\frac{f_{\mathcal{T}}u^{\tau}\varepsilon_{\zeta e
\omega}}{32 \pi
f_{R}}\mathcal{T}^{\zeta(m)}_{\tau}-\frac{f_{L_{m}}u^{\tau}\varepsilon_{\zeta
e \omega}}{64 \pi
f_{R}}\mathcal{T}^{\zeta(m)}_{\tau}-\frac{u^{\tau}\varepsilon_{\zeta
e \omega}}{32 \pi
f_{R}}\nabla^{\zeta}\nabla_{\tau}f_{R}+\frac{u^{\tau}\varepsilon_{e
\upsilon \omega}}{16 \pi
f_{R}}f_{\mathcal{T}}\mathcal{T}^{\upsilon(m)}_{\tau}\\\nonumber&+\frac{u^{\tau}\varepsilon_{e
\upsilon \omega}}{32 \pi
f_{R}}f_{L_{m}}\mathcal{T}^{\upsilon(m)}_{\tau}+\frac{u^{\tau}\varepsilon_{e
\upsilon \omega}}{32 \pi
f_{R}}\nabla^{\upsilon}\nabla_{\tau}f_{R}+\frac{f_{\mathcal{T}}\varepsilon_{e
\upsilon \omega}}{16 \pi
f_{R}}\frac{\partial^{2}L_{m}}{\partial
g^{\gamma}_{\tau}\partial g_{\mu \nu}}g^{\mu
\nu}u^{\tau}-\frac{u^{\tau}\varepsilon_{e \zeta \omega}g^{\mu
\nu}}{16 \pi f_{R}}\frac{\partial^{2}L_{m}}{\partial
g^{\zeta}_{\tau}\partial g_{\mu \nu}},
\\\nonumber
F&=-\frac{f_{\mathcal{T}}\mathcal{T}^{\mu(m)}_{\zeta}}{64
\pi f_{R}}\varepsilon^{\lambda\zeta\eta}\varepsilon_{\mu \lambda
\eta}+\frac{\Box f_{R}}{64 \pi f_{R}}\varepsilon^{\lambda \zeta
\eta}\varepsilon_{\mu \lambda
\eta}-\frac{\nabla^{\mu}\nabla_{\zeta}f_{R}}{64 \pi
f_{R}}\varepsilon^{\lambda \zeta
\eta}\varepsilon_{\mu\lambda\eta}-\frac{f}{128 \pi
f_{R}}\varepsilon^{\lambda \zeta \eta} \varepsilon_{\mu \lambda
\eta}\\\nonumber&-\frac{f_{\mathcal{T}}g^{\alpha \beta}}{32 \pi
f_{R}}\frac{\partial^{2}L_{m}}{\partial
g^{\mu}_{\zeta}\partial g_{\alpha\beta}}\varepsilon^{\lambda\zeta\eta}\varepsilon_{\mu\lambda\eta}-\frac{f_{\mathcal{T}}\mathcal{T}^{\vartheta(m)}_{\lambda}}{64
\pi
f_{R}}\varepsilon^{\lambda\zeta\eta}\varepsilon_{\zeta\vartheta\eta}-\frac{f_{L_{m}}\mathcal{T}^{\vartheta(m)}_{\lambda}}{128
\pi
f_{R}}\varepsilon^{\lambda\zeta\eta}\varepsilon_{\zeta\vartheta\eta}-\frac{\nabla^{\vartheta}\nabla_{\lambda}f_{R}}{64
\pi
f_{R}}\varepsilon^{\lambda\zeta\eta}\varepsilon_{\zeta\vartheta\eta}\\\nonumber&
-\frac{f_{\mathcal{T}}g^{\alpha \beta}}{32 \pi
f_{R}}\frac{\partial^{2}L_{m}}{\partial
g^{\vartheta}_{\lambda}\partial g_{\alpha \beta}}\varepsilon^{\lambda \zeta
\eta}\varepsilon_{\zeta \vartheta \eta},
\\\nonumber
D&=\frac{\alpha f_{\mathcal{T}}}{8 \pi f_{R}}+\frac{\alpha
f_{L_{m}}}{16 \pi f_{R}}-\frac{\Box f_{R}}{2 \pi f_{R}}
+\frac{\nabla ^{\lambda}\nabla_{\lambda}f_{R}}{16 \pi f_{R}}+\frac{f}{4
\pi f_{R}}-\frac{R f_{R}}{4 \pi
f_{R}}-\frac{f_{\mathcal{T}}L_{m}}{\pi
f_{R}}-\frac{L_{m}}{2 \pi}+\frac{f_{\mathcal{T}}g^{\alpha
\beta}}{8 \pi f_{R}}\\\nonumber&\times
\frac{g^{\lambda}_{\mu}\partial^{2}L_{m}}{\partial
g^{\mu}_{\lambda}\partial g_{\alpha
\beta}}-\frac{u^{\eta}u^{\zeta}f_{\mathcal{T}}}{8 \pi
f_{R}}\mathcal{T}^{(m)}_{\eta
\zeta}-\frac{u^{\eta}u^{\zeta}f_{L_{m}}}{16 \pi
f_{R}}\mathcal{T}^{(m)}_{\eta
\zeta}+\frac{u^{\eta}u^{\zeta}}{16 \pi f_{R}}g_{\eta
\zeta}\Box f_{R}-\frac{u^{\eta}u^{\zeta}}{16 \pi
f_{R}}\nabla_{\eta}\nabla_{\zeta}f_{R}-\frac{u^{\eta}}{64
\pi}\\\nonumber&\times \frac{u^{\zeta}}{ f_{R}}g_{\eta \zeta}
+\frac{u_{\lambda}u^{\zeta}\delta^{\lambda}_{\zeta}R f_{R}}{32 \pi
f_{R}}+\frac{u_{\lambda}u^{\zeta}\delta^{\lambda}_{\zeta}}{8 \pi
f_{R}}L_{m}f_{\mathcal{T}}+\frac{u^{\eta}u^{\zeta}g_{\eta
\zeta}}{16 \pi
f_{R}}L_{m}f_{{L}_{m}}-\frac{u^{\eta}u^{\zeta}}{8
\pi f_{R}}\frac{\partial^{2}L_{m}}{\partial
g^{\mu}_{\zeta}\partial g_{\alpha \beta}}g^{\alpha \beta}g_{\eta
\mu}f_{\mathcal{T}}\\\nonumber&-\frac{u_{\vartheta}u^{\lambda}\mathcal{T}^{\vartheta
(m)}_{\lambda}}{8 \pi f_{R}}f_{\mathcal{T}}
-\frac{u_{\vartheta}u^{\lambda}\mathcal{T}^{\vartheta (m)}_{\lambda}}{16 \pi
f_{R}}f_{L_{m}}+\frac{\Box f_{R}}{16 \pi f_{R}}
-\frac{u_{\vartheta}u^{\lambda}}{16 \pi
f_{R}}\nabla^{\vartheta}\nabla_{\lambda}f_{R}-\frac{f}{32 \pi
f_{R}}+\frac{R f_{R}}{32 \pi f_{R}}
\\\nonumber&+\frac{f_{\mathcal{T}}}{8 \pi}
\frac{L_{m}}{f_{R}}+\frac{f_{L_{m}}L_{m}}{16
\pi f_{R}}-\frac{u_{\vartheta}u^{\lambda}}{8 \pi
f_{R}}\frac{\partial^{2}L_{m}}{\partial
g^{\vartheta}_{\lambda}\partial g_{\alpha \beta}}g^{\alpha
\beta}f_{\mathcal{T}}+\frac{u_{\vartheta}u^{\zeta}\mathcal{T}^{\vartheta
(m)}_{\zeta}}{4 \pi f_{R}}f_{\mathcal{T}}
+\frac{u_{\vartheta}u^{\zeta}\mathcal{T}^{\vartheta (m)}_{\zeta}}{8 \pi
f_{R}}f_{L_{m}}+\frac{u_{\vartheta}u^{\zeta}}{4 \pi
f_{R}}\\\nonumber& \times \nabla^{\vartheta}\nabla_{\zeta}
f_{R}+\frac{u_{\vartheta}u^{\zeta}f_{\mathcal{T}}}{4 \pi f_{R}}
\frac{\partial^{2}L_{m}}{\partial
g^{\vartheta}_{\zeta}\partial g_{\alpha \beta}}g^{\alpha \beta}.
\end{align}

\end{document}